\documentclass[a4paper,fleqn,usenatbib]{mnras}

\usepackage{newtxtext,newtxmath}

\usepackage[T1]{fontenc}
\usepackage{ae,aecompl}

\usepackage{graphicx}	
\usepackage{amsmath}	
\usepackage{amssymb}	

\title[]{The GALAH Survey: Temporal Chemical Enrichment of the Galactic Disk}

\author[Lin et al.]{
Jane Lin$^{1,2}$\thanks{E-mail: jane.lin@anu.edu.au},
Martin Asplund$^{1,2}$,
Yuan-Sen Ting$^{3,4,5,6}$
Luca Casagrande$^{1,2}$,
Sven Buder$^{7}$,  
\newauthor
Joss Bland-Hawthorn$^{2,8}$,
Andrew R. Casey$^{9}$,
Gayandhi M. De Silva$^{10}$,
Valentina D'Orazi$^{11}$,
\newauthor
Ken C. Freeman$^{1,2}$,
Janez Kos$^{12}$,
K Lind$^{7,13}$,
Sarah L. Martell$^{2,14}$,
Sanjib Sharma$^{8,2}$,
\newauthor
Jeffrey D. Simpson$^{14}$,
Toma\v{z} Zwitter$^{12}$,
Daniel B. Zucker$^{10}$,
Ivan Minchev$^{15}$,
Klemen \v{C}otar$^{12}$,
\newauthor
Michael Hayden$^{8,2}$,
Jonti Horner$^{16}$,
Geraint F. Lewis$^{8}$
Thomas Nordlander $^{1,2}$
\newauthor
Rosemary F. G. Wyse $^{17}$
and
Maru{\v s}a {\v Z}erjal $^{1,2}$
\\
$^{1}$Research School of Astronomy and Astrophysics, The Australian National University, Canberra, ACT 2611, Australia\\
$^{2}$ ARC Centre of Excellence for All Sky Astrophysics in 3 Dimensions (ASTRO 3D) \\
$^{3}$Institute for Advanced Study, Princeton, NJ 08540, USA\\
$^{4}$ Department of Astrophysical Sciences, Princeton University, Princeton, NJ 08544, USA\\
$^{5}$Observatories of the Carnegie Institution of Washington, 813 Santa Barbara Street, Pasadena, CA 91101, USA\\
$^{6}$Hubble Fellow\\
$^{7}$Max-Planck-Institut f\"ur Astronomie, K\"onigstuhl 17, D-69117 Heidelberg, Germany\\
$^{8}$Sydney Institute for Astronomy, School of Physics, University of Sydney, NSW 2006, Australia\\
$^{9}$School of Physics and Astronomy, Monash University, Clayton 3800, Victoria, Australia\\
$^{10}$Department of Physics \& Astronomy, Macquarie University, Sydney, NSW 2109, Australia\\
$^{11}$INAF -Osservatorio Astronomico di Padova\\
$^{12}$Faculty of Mathematics and Physics, University of Ljubljana, Jadranska 19, 1000 Ljubljana, Slovenia\\
$^{13}$Observational Astrophysics, Department of Physics and Astronomy, Uppsala University, Box 516, SE-751 20 Uppsala, Sweden\\
$^{14}$School of Physics, University of New South Wales, Sydney, NSW 2052, Australia\\
$^{15}$Leibniz-Institut ̈f\"ur Astrophysik Potsdam (AIP), An der Sternwarte 16, 14482 Potsdam, Germany\\
$^{16}$Centre for Astrophysics, University of Southern Queensland, Toowoomba, Queensland 4350, Australia\\
$^{17}$ Dept of Physics and Astronomy, Johns Hopkins University, Baltimore, MD 21218,
USA
}

\date{Accepted XXX. Received YYY; in original form ZZZ}

\pubyear{2015}

\begin{document}
\label{firstpage}
\pagerange{\pageref{firstpage}--\pageref{lastpage}}
\maketitle

\begin{abstract}

We present isochrone ages and initial bulk metallicities ($\rm [Fe/H]_{bulk}$, by accounting for diffusion) of 163,722 stars from the GALAH Data Release 2, mainly composed of main sequence turn-off stars and subgiants ($\rm 7000 K>T_{eff}>4000 K$\ and $\rm log g>3$\,dex).
The local age-metallicity relationship (AMR) is nearly flat but with significant scatter at all ages; the scatter is even higher when considering the observed surface abundances. After correcting for selection effects, the AMR appear to 
have intrinsic structures indicative of two star formation events, which we speculate are connected to the thin and thick disks in the solar neighborhood. We also present abundance ratio trends for 16 elements as a function of age, across different $\rm [Fe/H]_{bulk}$ bins. In general, we find the trends in terms of [X/Fe] vs age from our far larger sample to be compatible with studies based on small ($\sim$ 100 stars) samples of solar twins but we now extend it to both sub- and super-solar metallicities.
The $\alpha$-elements show differing behaviour: the hydrostatic $\alpha$-elements O and Mg show a steady decline  with time for all metallicities while the explosive $\alpha$-elements Si, Ca and Ti are nearly constant during the thin disk epoch (ages $\la 12$\,Gyr).
The s-process elements Y and Ba show increasing [X/Fe] with time while the r-process element Eu have the opposite trend, thus favouring a primary production from sources with a short time-delay such as core-collapse supernovae over long-delay events such as neutron star mergers.

\end{abstract}




\section{Introduction}
The Milky Way occupies a special place in understanding galactic evolution. Unlike other galaxies, within the Milky Way we are able to resolve individual stars at great precision and disentangle their chemodynamical relationships. Galactic archaeology \citep{freeman2002new} aims to use present day stellar abundances as fossils to trace the formation and evolutionary history of the Milky Way. 
Combining stellar abundances with ages allows us to track the chemical enrichment of the Milky Way and thus provide crucial insights into the star formation, assembly (accretion/merger/outflow), and dynamical history of the Galaxy and its stars and gas. This is possible because different elements are produced through a variety of nucleosynthetic production channels, each with its own characteristic timescales and physics \citep[e.g.,][]{travaglio2004galactic,frebel2005nucleosynthetic,heger2002nucleosynthetic}. Studying stars at different ages allows investigations into behaviours of elements during earlier stages of the Milky Way and provides the vital temporal axis in galaxy evolution. Indeed, the absence of age information can result in misinterpretations of the Milky Way formation and evolution  \citep[see][and the examples therein]{2019MNRAS.487.3946M} . 

Historically, the determination of stellar ages has been challenging, as age is not a directly observable quantity. Our understanding of stellar ages strictly relies on stellar evolutionary models, unlike other parameters such as $\rm [\alpha / Fe]$ which can be directly inferred. Instead, metallicity and $\rm [\alpha/Fe]$ have traditionally been used as proxies for age \citep[e.g.][]{tinsley1979stellar,edvardsson1993chemical,ryan1996extremely,prochaska2000galactic, 2012ApJ...753..148B,2013MNRAS.434..652T}. The limitation in these age proxies is the significant amount of scatter in abundances at any given age bin \cite[e.g.,][]{2018MNRAS.481.1645M,2017MNRAS.471.3057M}.

Most methods used today require stellar evolutionary models one way or another and still have large uncertainties. With advances in stellar modelling and precise distances \citep[e.g.,][]{perryman1997hipparcos}, it is only relatively recently that deriving ages for large number of field stars has become possible. The most common way of deriving stellar ages is by isochrone fitting \citep[e.g.,][]{jorgensen2005determination, yi2001toward}. Isochrones are paths of equal time on the  Hertzsprung-Russell Diagram (HRD), computed from stellar evolutionary tracks, \citep[e.g.,][]{bressan2012parsec,dotter2008dartmouth,choi2016mesa}, taking into account various stellar physics factors. The stellar models are relatively robust (especially for solar-like stars), as the underlying physics is by now better understood. Main sequence turn-off and subgiants make ideal isochrone fitting candidates because of the large stellar parameter to age sensitivity in this region of the HRD. In addition to isochrone fitting, asteroseismology allows us to derive masses and ages for main-sequence and giant stars displaying solar-like oscillations \citep[e.g.,][]{garcia2014rotation,chaplin2011ensemble,metcalfe2010precise}.  Space missions like \textit{CoRoT} \citep{auvergne2009corot} and \textit{Kepler}/\textit{K2} \citep{2014PASP..126..398H} have been instrumental in providing high precision time-series photometry for determination of asteroseismic parameters. Recent and future missions such as \textit{TESS} \citep{ricker2014transiting} and \textit{PLATO} \citep{rauer2016plato} will also play integral roles in asteroseismology. 

Multiple dating methods exist for other evolutionary stages. Ages of young stars are sensitive to rotation, as loss of birth angular momentum induces uniformity in rotation rates. Over time, magnetic braking causes rotation to  slow down. Using star spots, rotation velocities can be measured photometrically with light curves. Ages are then inferred by comparing them to cluster calibrated relations \citep[e.g.,][]{2007ApJ...669.1167B,2008ApJ...687.1264M,2013ApJ...776...67V}. Similarly, the loss of angular momentum makes the star less magnetically active. This decrease in activity and hence age can be inferred from certain emission spectral features such as Ca II H and K lines \citep[e.g.,][]{2008ApJ...687.1264M,2013ApJ...776..127Z,2017ApJ...835...61Z}. For red giants, C and N abundances are mass sensitive, as the first dredge-up changes their ratio in post main sequence evolution \citep{2016AN....337..805S}. Ages are then inferred from masses using isochrones \citep[e.g.,][]{2016MNRAS.456.3655M}. \citet{2010ARA&A..48..581S} provides a detailed break down of various dating methods for different evolutionary stages. 

More recent studies have derived robust ages for solar-like field stars, coupled with precise stellar abundances \citep[e.g.,][]{2015A&A...579A..52N,2016A&A...593A.125S,2016A&A...590A..32T, 2018arXiv180202576B}. However they often have only relatively small samples of stars  \citep[up to $\sim 10^3$ stars, e.g.,][]{2012A&A...545A..32A,2013A&A...560A.109H}, and usually do not account for their heterogeneous selection effects. Only very recently has isochrone ages of large samples of stars ($\sim 10^5$ stars) become available \citep[e.g.,][]{2018A&A...618A..54M,2018MNRAS.476.2556Q,2018MNRAS.tmp.2388S}, this is coupled with the data releases of large scale spectroscopic surveys such as GALAH \citep{2015MNRAS.449.2604D, 2018MNRAS.478.4513B}, LAMOST \citep{luo2015first}, APOGEE \citep{holtzman2015abundances}, Gaia-ESO \citep{gilmore2012gaia} and parallax information from Gaia \citep{2018A&A...616A...1G}. The large number of stars covered by these surveys make separation of different galactic components and populations more achievable. 

The Solar neighborhood contains stars with a wide range of ages and birth locations \citep{2018MNRAS.481.1645M,2018ApJ...865...96F}. The substantial number of turn-off and subgiants within the GALAH survey data set is  an ideal sample for isochrone fitting and thus allows us to investigate the age-abundance trends for an unprecedented number of stars, reconstructing the chemical enrichment history of the Galaxy. 

This paper is structured as follows: in Section~\ref{sec:methods} we introduce the GALAH sample, our isochrone fitting method and address the GALAH selection effect. Section~\ref{sec:amr} discusses the age metallicity relationship in the sample. The age-abundance trends are presented in Section~\ref{sec:trends}, followed by conclusion and discussions. 

\section{Methods}\label{sec:methods}

\subsection{GALAH DR2}
The GALAH survey is specifically designed to have a relatively simple selection function, targeting stars with apparent magnitude $12 \rm \leq	V \leq 14$, where V is estimated from 2MASS photometry \citep{skrutskie2006two} using the following relation
\[    \rm V=K+2(J-K+0.14)+0.382e^{(J-K-0.2)/0.5} \]

\noindent 
without any colour cuts. Having a simple selection function allows its effects to be well accounted for, which is vital for galactic inference. The observations are performed using the HERMES multi-object spectrograph \citep{sheinis2014first} on the 3.9\,m Anglo-Australian Telescope at Siding Spring Observatory, Australia, enabling $\sim 360$ stars to be observed simultaneously to acquire high-resolution ($R=\lambda/\Delta\lambda \approx 28,000$) spectra in four wavelength regions.
The observed spectra are reduced using an IRAF \citep{tody1986iraf} pipeline, as described in \citet{2017MNRAS.464.1259K}. 

GALAH mainly targets the thin disk with $|b|>10^{\circ}$ and $-80^{\circ} <\delta<+10^{\circ} $ \citep{2015MNRAS.449.2604D}. A detailed description of stellar parameter and abundance determination for the GALAH survey is presented in \citet{2018MNRAS.478.4513B} and here we only briefly summarise the procedure. Stellar parameters ($\rm T_{eff}, log g, [Fe/H], V_{sin i}, radial\ velocity$) and  $\rm \alpha$-abundances were derived using a combination of spectral synthesis and data-driven models. For spectral synthesis, selection was made for a sample of 10,605 high quality training stars encompassing the parameter space covered in the survey, as well as benchmark stars with stellar parameters obtained independently of spectroscopy \citep{2015A&A...582A..49H}. Spectroscopy Made Easy \citep[SME,][]{valenti1996spectroscopy,piskunov2017spectroscopy} was used to fit the synthetic spectra to the observed. Synthetic spectra were generated using  MARCS model atmospheres \citep[][]{2008A&A...486..951G}, with non-LTE corrections for O, Na, Mg, Al, Si and Fe \citep[][]{2012MNRAS.427...50L}. Stellar parameters were then propagated from the training set to the rest of GALAH sample using The Cannon \citep{2015ApJ...808...16N}. It builds data-driven spectral approximations by modelling how the flux varies as a function of stellar labels, assuming a quadratic function. This empirical spectral model was then used to fit for other stars through $\rm \chi^2$ minimization. Our GALAH sample also includes stars from the TESS-HERMES \citep{2018MNRAS.473.2004S} and K2-HERMES \citep{2018AJ....155...84W} programs. 

We have restricted our sample to mainly the turn-off region ($\rm 7000K>T_{eff}>4000K$\ and $\rm log g>3$\,dex), as isochrone ages are most robust in this area of HRD. On the other hand, regions like red giant branch (RGB) and main-sequence (MS) have stellar parameters which are less sensitive to age, as they have very high isochrone densities. In addition to ages, the stellar parameters themselves are more uncertain in regions outside of our selection box due to the paucity of training stars at the hottest and coolest ends of the temperature scale. Hence this study is focused primarily in the sub-giant and turn-off regions. 

\subsection{Age determination}\label{sec:aged}

Stellar ages and masses are determined by Bayesian isochrone fitting, as described in \citet{2018MNRAS.477.2966L}. In brief, we employ a grid of MIST isochrones \citep{choi2016mesa}, which spans $5.0-10.3$ in log(age\,yr), $0.1-300$\,$\rm M_{\odot}$ in initial mass and $\rm -2.0 \leq [Fe/H]_{bulk} \leq + 0.5$ in initial bulk metallicity. We make the important distinction between the initial bulk composition of the stellar models ($\rm [Fe/H]_{bulk}$) and the present day observed surface metallicity ($\rm [Fe/H]_{surf}$). $\rm [Fe/H]_{bulk}$ is treated as an input parameter (the posterior of which we will sample), whereas $\rm [Fe/H]_{surf}$ is treated as a model prediction (which we will compare to observed metallicity), it is affected by stellar mixing, gravitational settling and atomic diffusion \citep{1994ApJ...421..828T}. These effects are most prominent near the MS and turn-off regions on the HRD. The difference in metallicities can be as large as 0.5\,dex, for a young, metal-poor star and around 0.05\,dex for a star at solar metallicity and age. If unaccounted for, this can potentially bias age estimates by as much as 20\% systematically \citep{dotter2017influence}. On the other hand, taking into account of this differentiation allows us to compare stellar abundances at different evolutionary stages.

In our posterior ($\rm p_1$), we sample for age ($\tau$), evolutionary state ($\rm EEP$, defined as equivalent evolutionary stages under the MIST scheme), $\rm [Fe/H]_{bulk}$, parallax ($\rm \bar{\omega}_{sample}$), K-band extinction ($\rm A_K$):

\begin{multline*}
p_1(\tau,\rm{EEP},\rm{[Fe/H]_{bulk},A_K, \bar{\omega}_{sample}} |T_{\rm eff}, {\rm log}g, {\rm [Fe/H]_{surf}}, {\rm m_{K}}, \bar{\omega}_{model})\\
\propto  \\
\mathcal{L}( O|\tau,\rm{EEP},\rm{[Fe/H]_{bulk},A_K, \bar{\omega}_{sample}} ) \pi
\end{multline*}

In the likelihood ($\mathcal{L}$), we compare model predictions ($ \{ \rm T_{eff}, log g, [Fe/H]_{surf}, \bar{\omega}_{model}, m_K \}$) with their corresponding observed values $O$: $ \{ \rm T_{eff}, log g, [Fe/H]_{surf},m_K, \bar{\omega}_{Gaia} \}$. The first three are spectroscopically determined, $\rm \bar{\omega}_{Gaia} $ is the Gaia parallax \citep{2018A&A...616A...1G} and $\rm m_K$ is apparent K magnitude from 2MASS \citep{cutri20032mass}. We chose to make the distinction between $\rm \bar{\omega}_{model}$ and $\rm \bar{\omega}_{sample}$ to keep the notation consistent. Similarly, by comparing $\rm \bar{\omega}_{model}$ to $\rm \bar{\omega}_{Gaia}$, we are effectively drawing from $\rm \bar{\omega}_{Gaia }$ prior.  We decide to sample EEP instead of mass because even though both EEP and mass are one to one mapped to age, but the observables vary more smoothly with respect to EEP, hence pragmatically easier to interpolate. 

We adopt a non-informative flat prior ($\pi$) in extinction:

\[   
\pi(A_K) = 
     \begin{cases}
       \text{1} &\quad\text{for}\ A_S > A_K > 0\\ 
       \text{$0$} &\quad\text{else}. \\
     \end{cases}
\]

\noindent
Here $\rm A_S$ is taken from \citet{schlegel1998maps}. We furthermore force the sampled parallax ($\rm \bar{\omega}_{sample}$) to be positive. The posterior is sampled using  Markov chain Monte Carlo (MCMC) via the Python package \textit{emcee} \citep{foreman2013emcee}. 

\begin{figure}
\includegraphics[width=9.2cm]{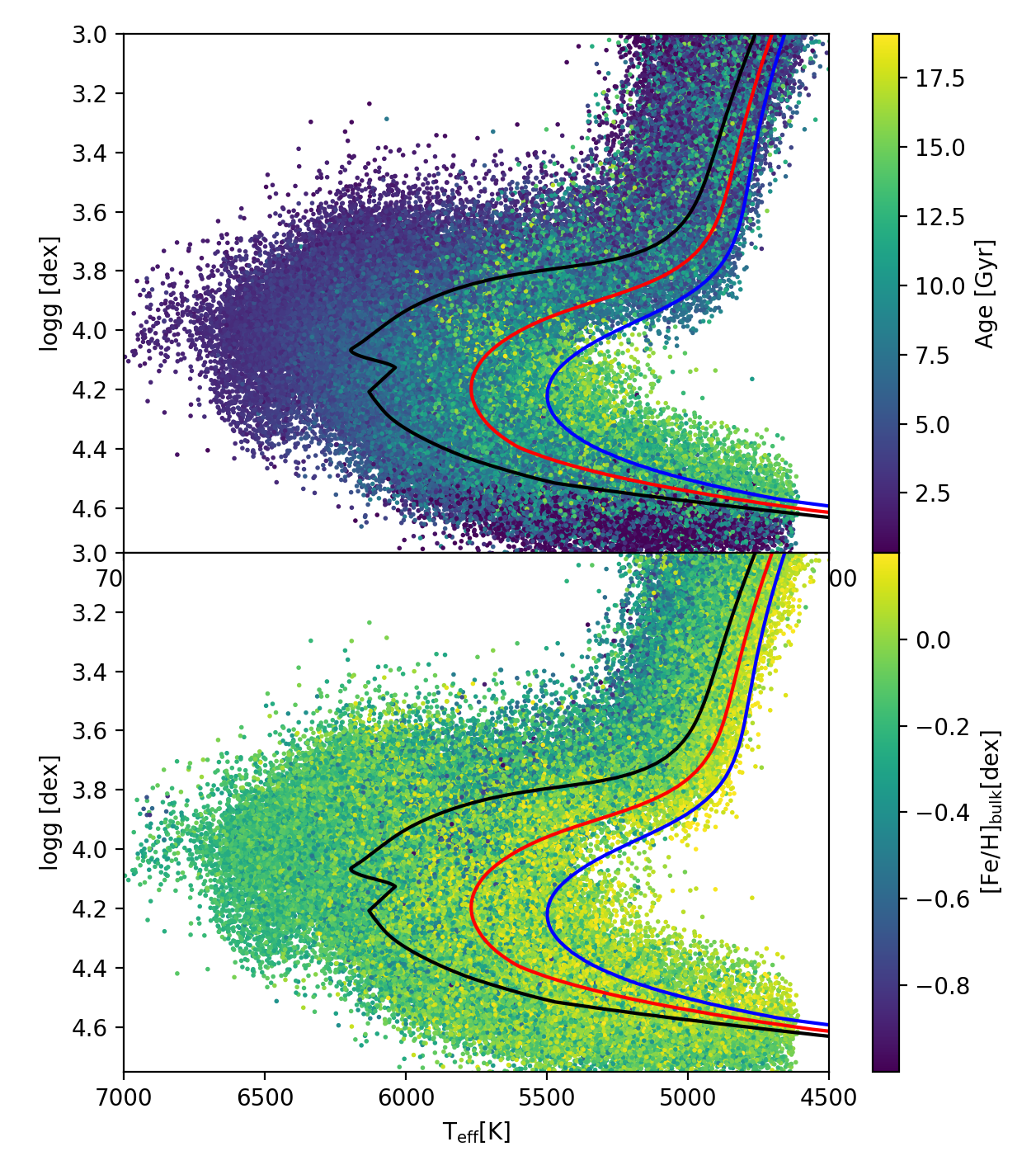}
\caption{Top panel: the complete sample (163,722 stars) colour coded in age, over plotted with solar metallicity isochrones, for ages 5 (black), 10 (red) and 16 (blue)\,Gyr. Bottom panel: sample colour coded in bulk metallcity (only stars with bulk metallicities greater than -1\,dex are shown here). }
\label{fig:hrs}
\end{figure}

We present ages and evolutionary stages of 163,722 stars (out of which, 126,152 stars have relative age uncertainty lesser than 30\%). Figure~\ref{fig:hrs} shows our complete sample colour coded in age and bulk metallicity, with three solar metallicity isochrones over plotted.

\begin{figure}
\includegraphics[width=9.2cm]{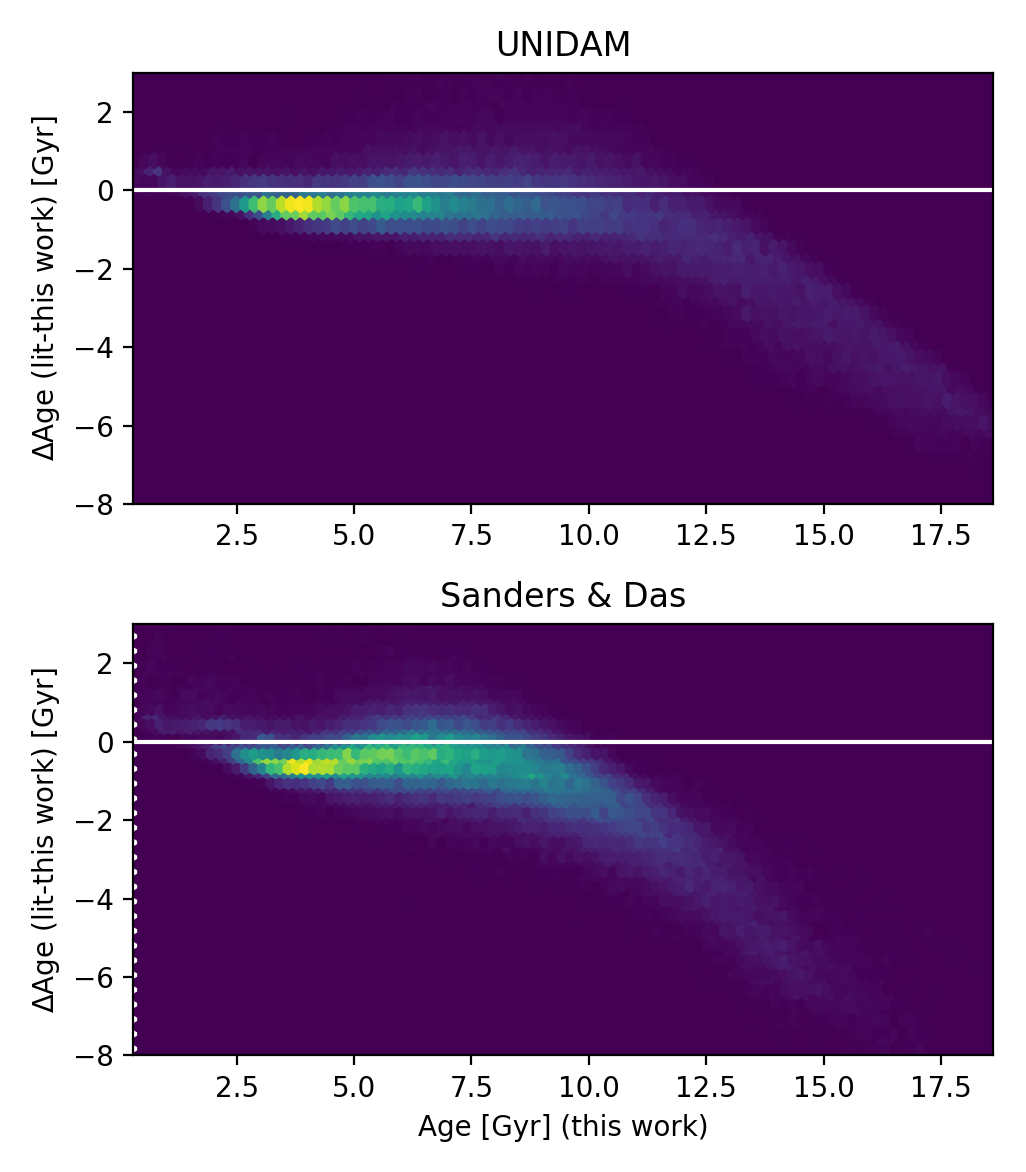}
\caption{Differences in age (literature-this work) for UNIDAM (top panel) and \citet{2018MNRAS.tmp.2388S} (bottom panel). Only stars with uncertainty less than 30\% are plotted.}
\label{fig:comp}
\end{figure}

Overall, our ages are comparable with literature samples UNIDAM~\citep{2017A&A...604A.108M} and \citet{2018MNRAS.tmp.2388S}. Selecting only stars with relative age uncertainties lesser than 30\% and passing quality cuts, we have 70,746 stars in common with UNIDAM and 94,766 with \citet{2018MNRAS.tmp.2388S}. Figure~\ref{fig:comp} shows the age comparison between these two literature values and this work. Our ages are slightly higher ($\sim$0.5-1\,Gyr, most likely due to our choice of isochrone grids) for stars below 10\,Gyr and significantly higher for older stars. 

Multiple factors contribute to our expanded age scale. Firstly, the current MIST isochrones have no alpha enhancement at subsolar metallicities, this leads to ages being systematically older under the MIST scheme. We test this alpha enhancement effect by modifying metallicity, as prescribed by \citet{1993ApJ...414..580S} and calculate ages for our GALAH sample. As expected, we find the non-metallicity modified sample gives consistently older ages, with the differences in age between the two samples being the greatest at higher ages (with the mean absolute difference in age being 0.3\,Gyr for stars below 10\,Gyr and increasing to 0.9\,Gyr for stars older than 10\,Gyr). This effect contributes to the divergence in older ages observed between our sample and the two literature samples in Figure~\ref{fig:comp}.

Uncertainties and offsets in GALAH stellar parameters also contribute to older ages in some stars. To test this, we generate a sample of mock stars at 12.6\,Gyr using MIST and perturb their stellar parameters by typical GALAH uncertainties while also assuming GALAH-like uncertainties for the perturbed parameters. The resulting age distribution exhibits a range of plausible ages, including to up to 20\,Gyr. \citet{2019arXiv190607489R} also observes that stellar parameters with relative uncertainties typical of present day surveys can result in ages up to 20\,Gyr. This is akin to negative Gaia parallaxes observed for some non-GALAH stars- they still convey information on the lower limits of the corresponding distances, similarly, a large age still conveys information that the star is likely a very old one. Lastly, MIST adopts \citet{asplund2009chemical} solar abundances, which are significantly lower than the canonical values \citep[e.g.][]{1989GeCoA..53..197A,1998SSRv...85..161G} typically used in other popular isochrone grids. 

The interplay of these effects leads to an expanded age scale (up to 20\,Gyr), whereas both literature samples are computed using alpha enhanced PARSEC isochrones \citep{2012MNRAS.427..127B} with a maximum cutoff age of 13.5\,Gyr for UNIDAM and 12.6\,Gyr for \citet{2018MNRAS.tmp.2388S}. The 12.6\,Gyr cutoff means that potentially there will be a pileup of stars near terminal ages, as observational uncertainties can cause certain stars to favor older ages. Due to our expanded age scale, this paper is more focused on the relative chronology of events, rather than absolute ages.

\subsection{Observational selection effects}\label{sec:selection}
 Many methods exist to account for biases introduced by different selection cuts, including involving population synthesis and galactic priors   \citep[e.g.,][]{2014A&A...565A..89B, 2016ApJ...817...40F,2019arXiv190210485E}. Indeed, to fully explore the selection effects of the GALAH survey warrants a separate paper in itself. Here we apply a simple and least model dependent way to correct for selection effects, as described in \citet{2016MNRAS.455..987C}. 
 
 Firstly a grid in $\rm [Fe/H]_{bulk}$ and age is constructed using MIST isochrones. At each grid point, we populate the isochrone using $10^6$ stars, with masses distributed according to \citet{2001MNRAS.322..231K} initial mass function (IMF). Once an isochrone is populated with the IMF, it accounts for different evolutionary speeds of various masses and hence the probability of observing different evolutionary stages. We then project the absolute $V$ magnitude of these stars to different distances ranging from 10 to 7910\,pc at 100\,pc intervals, thus obtaining a set of observed magnitudes to which we can apply the same selection function as the data. We apply two cuts: the first is the $12 \leq	V \leq 14$ magnitude cut from the main GALAH target selection, the second are $\rm 7000>T_{eff}>4000$\,K and $\rm log g>3$\,dex, introduced by the lack of training stars outside of this range for the GALAH spectroscopic analysis pipeline.
 
\begin{figure}
\includegraphics[width=9.2cm]{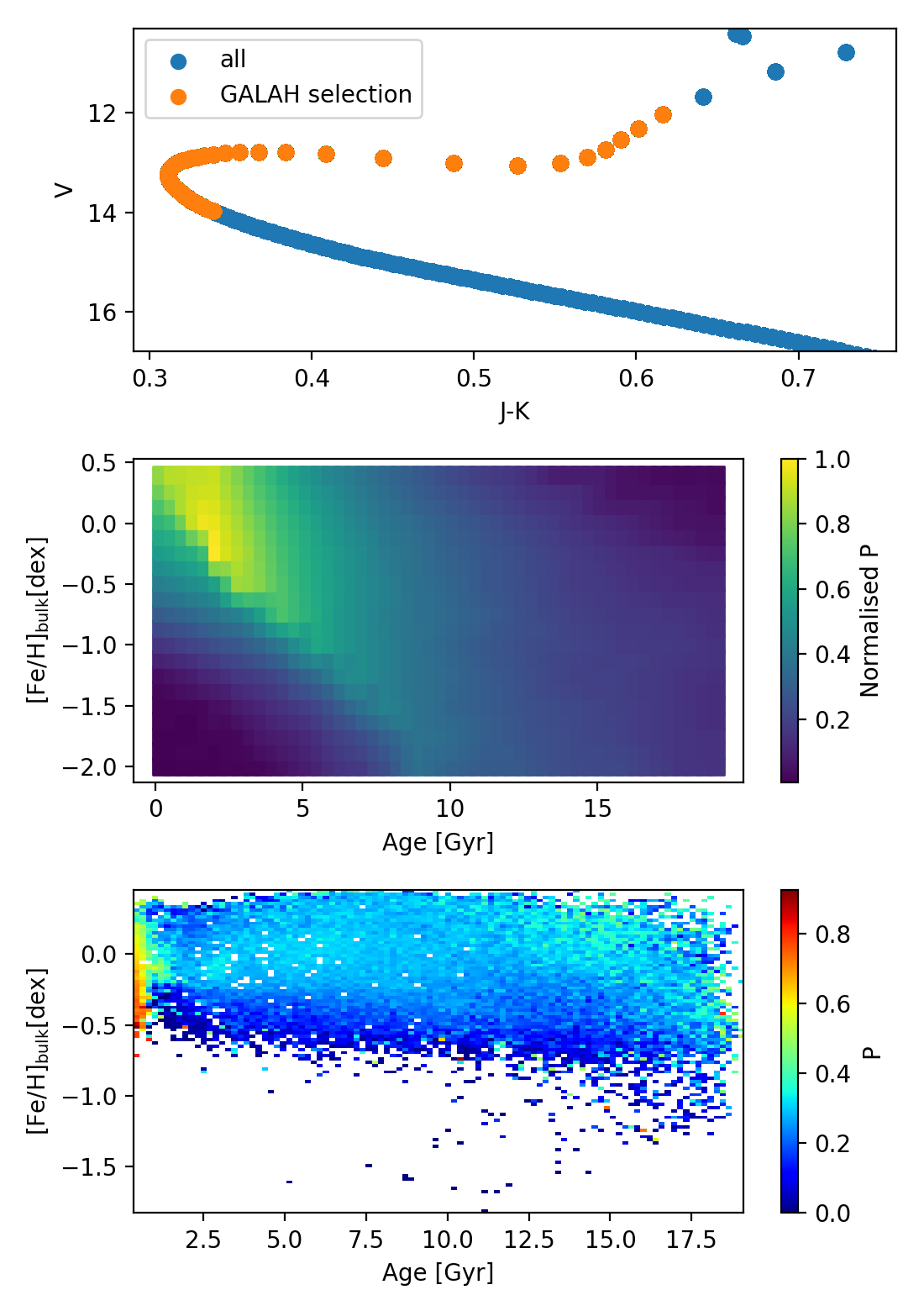}
\caption{Top panel: synthetic population at $\rm [Fe/H]_{bulk}=$ 0\,dex and 5.6\,Gyr, with apparent J-K and V magnitudes projected to 800\,pc. Due to the IMF, there are far more low mass stars compared to massive stars. Orange points represent the subsect of stars remaining after the GALAH selection cuts. Middle panel: scaled probability as a function of age and $\rm [Fe/H]_{bulk}$, at a distance of 800\,pc. Young, metal-poor stars (too bright) and old, metal-rich stars (too dim) are under sampled as a result of our selection effect. Bottom panel: probability map for our complete sample (163,722 stars), averaged over 100 bins in either directions. }
\label{fig:selection}
\end{figure}

 This allows us to calculate the observing probability (P): the relative fraction of stars which remains after these cuts for any combination of \{$\rm [Fe/H]_{bulk}$, age and distance\}. Top  panel of Figure~\ref{fig:selection} shows an example synthetic population generated at $\rm [Fe/H]_{bulk} = 0.0$\,dex and age 5.6\,Gyr, with apparent J-K and V magnitudes projected to a distance of 800\,pc. The orange points correspond to fraction of stars remaining after the GALAH survey selection effect.
 
 Finally we use \texttt{scipy.LinearNDInterpolator} to interpolate within the grid such that we can calculate P for any combination of \{$\rm [Fe/H]_{bulk}$, age and distance\}. Middle panel of Figure~\ref{fig:selection}  shows an example of the probability distribution as a function of $\rm [Fe/H]_{bulk}$ and age, for the distance of 800\,pc. Extinction is assumed to be negligible when performing distance projections as GALAH avoids high extinction regions and 2MASS magnitudes minimise their effects. Bottom panel of Figure~\ref{fig:selection} shows the resulting probability map for our sample in terms of age and metallicity. 
 

As clear from Figure~\ref{fig:selection},
the GALAH target selection biases against stars with low bulk metallicity for all ages and favours very young stars ($\rm <$0.6\,Gyr). In particular, there is a region of extremely low P for metal-poor, relatively young stars (roughly between 0.5 to 2\,Gyr (this region is somewhat averaged out with the binning applied to the figure). To confirm the existence of this low P region, we build a similar P map of the UNIDAM AMR using this method, replacing $\rm [Fe/H]_{bulk}$ with $\rm [Fe/H]_{surf}$ and switching to BASTI \citep{pietrinferni2004large} isochrones, so that the age scales are comparable. We are able to reproduce the extremely low P regions of metal-poor, relatively young stars, as well as the over-all topography of the P map. Another feature of the P map is the very young and high P region at the beginning of the AMR. The UNIDAM sample has very few stars below 0.5\,Gyr and BASTI grid does not extend below 0.5\,Gyr, making it hard to replicate this high P young region ($\rm <$0.6\,Gyr). The existence of this young population is also observed by Miglio et al. (in prep).
 
 We also observe an area of low P between 12.5 and 16\,Gyr (near $\rm [Fe/H]_{bulk}= $-0.25 - -0.75\,dex), which mainly consists of old stars with low $\rm [Fe/H]_{bulk}$, they are likely to be thick disk stars. This low-P region indicates that the observed data preferentially undersample the thick disk stars, and hence the data could erase any age peak that is associated to the thick disk if the selection function is not properly taken into account.
 
In addition to observational selection effects, we also consider age uncertainties. For this study, we do not impose any cuts on the relative age uncertainty because it would predominantly affect the main sequence and red giant branch regions of the HR-diagram (where the age sensitivity is very low), hence introducing another layer of selection bias. We choose instead to sample the uncertainty distributions for each star. We assume uncertainties in age and $\rm [Fe/H]_{bulk}$
 can be represented as 2D Gaussians with $\rm \sigma= \sigma(\tau)$ and  $\rm \sigma([Fe/H]_{bulk})$ and draw from the distribution 100 times for each star. We calculate P for each draw and weight the point accordingly.

\section{The GALAH age-metallicity relationship} \label{sec:amr}

\begin{figure}
\includegraphics[width=9.2cm]{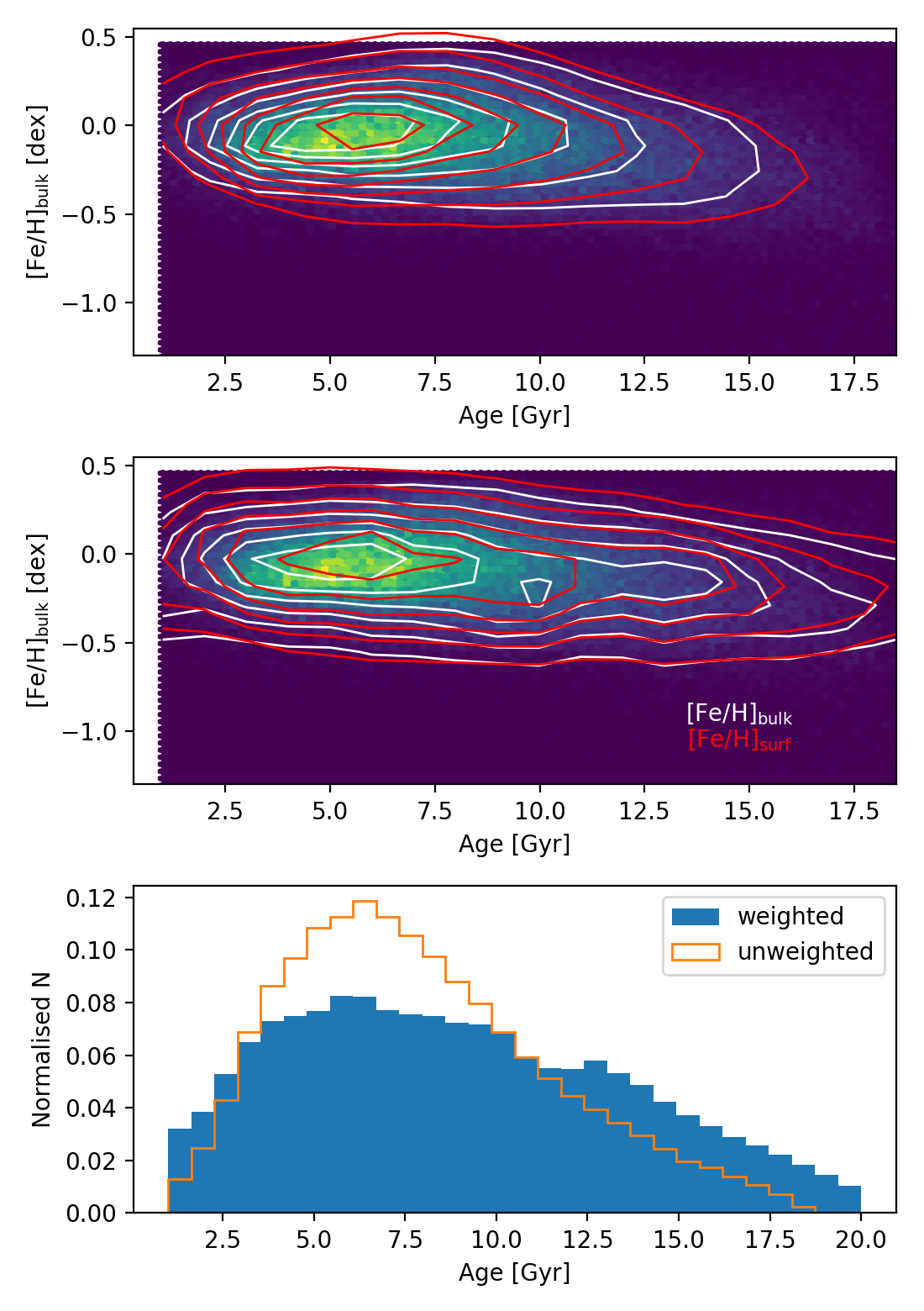}
\caption{Top panel: unweighted AMR with  $\rm [Fe/H]_{bulk}$ (white) and $\rm [Fe/H]_{surf}$ (red) contours, each binned over 17 bins in both directions. Middle panel: weighted AMR with both metallicity contours. Bottom panel: histogram of ages for both unweighted and weighed samples. }
\label{fig:amr}
\end{figure} 

Top panel of Figure~\ref{fig:amr}  shows the AMR contour without weighting by P. We have omitted stars younger than 1.5\,Gyr in the contour plots and age distribution. This is largely a cosmetic choice, as some of the aforementioned young, metal-poor stars have very low P, hence high weights, skewing the plots. The red and white contours represent bulk and surface metallicities respectively. We see a paucity of stars between 1.5 and 2.5\,Gyr (mainly absent in high temperatures), which is largely due to the data driven Cannon-based GALAH parameters having systematic offsets at high temperatures and are therefore discarded \citep[see Figure 6 of ][]{2019MNRAS.482.2770C}. Without weighting for selection effects, we observe an AMR which is commonly reported in literature: flat, with large scatter in metallicity for all ages before $\rm \sim$12\,Gyr, then a slight downward trend for the oldest stars  \citep[e.g.,][]{2011A&A...530A.138C,2014A&A...565A..89B,edvardsson1993chemical,feltzing2001solar}.  

The AMR changes shape once selection weights are applied (middle panel of  Figure~\ref{fig:amr}), becoming more elongated as older stars are generally selected against.
The elongated AMR have a downward trend for the highest ages, which could be the same trend of kinematically thick disk stars observed in \citet{2004A&A...421..969B}. Interestingly, there appears to be some kind of sub-structure in terms of age in the weighted AMR. This is more prominent in the weighted age distribution (bottom panel of Figure~\ref{fig:amr}). One possible explanation for this could be the manifestation of the local thin disk (the primary and broad  distribution peaking around 6\,Gyr, containing the bulk of our stars) and the chemical thick disk in the solar neighborhood (the smaller, secondary peak near 13\,Gyr). The chemical thick disk generally has a higher alpha elemental abundance relative to iron and is older in age compared to the thin disk. We remind the reader that our MIST-based isochrone ages are stretched due to the adopted compositions and isochrones, which leads to unrealistically high ages and also because of  potential mismatches in GALAH and MIST abundance scales- in reality the thick disk over density is more likely to correspond to 10-12\,Gyr as previously advocated \citep[e.g.,][]{2014A&A...562A..71B,2013A&A...560A.109H,2016ApJ...831..139M}. 

\begin{figure}
\includegraphics[width=9.2cm]{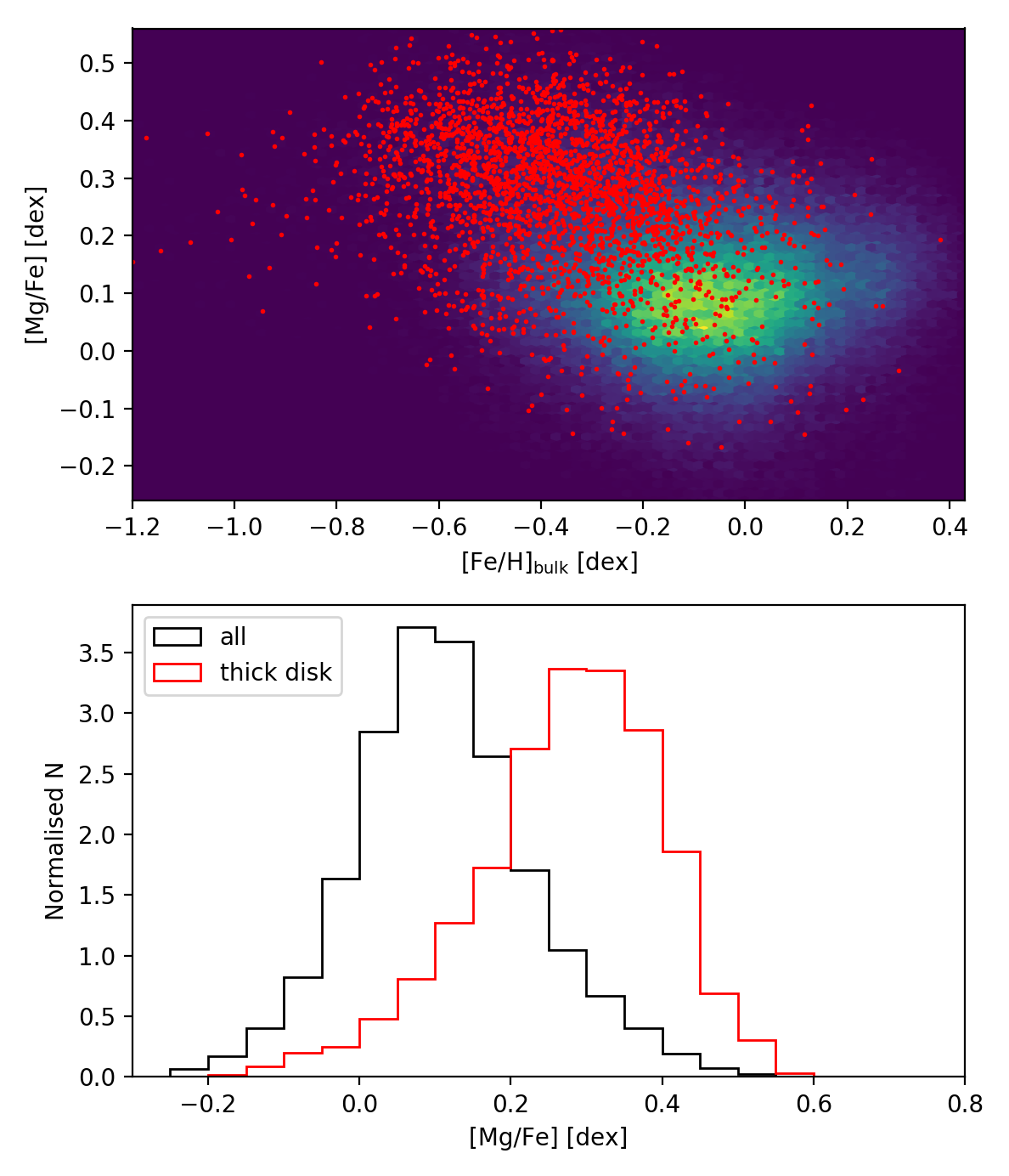}
\caption{Top panel: Iron and magnesium abundances of the thick disk sample. Red points are proposed thick disk stars selected to have ages between 12.5 and 16.0\,Gyr and P$<$0.1. Bottom panel:  normalised $\rm [Mg/Fe]$ distributions of all stars (black) and thick disk stars (red).}
\label{fig:mgfe}
\end{figure}

If confirmed, our observed AMR with multiple components is enticing and fits nicely with our current understanding of thin/thick disk formation as well as other observations. The dual nature of the age distribution would be consistent with two separate star formation events, one for each disk, as outlined in \citet{1998A&A...338..161F}, where the thin disk is formed continuously and the thick disk is formed more rapidly in one star-forming event, with a quenching in star formation between the two \citep{2016A&A...589A..66H}.  The recent discovery of the Gaia-Enceladus accretion  \citep[e.g.,][]{2018Natur.563...85H,2018MNRAS.478..611B,2018ApJ...863L..28M} could be a potential candidate associated with this sequence of events.  The thick disk population is not only older and more metal-poor, it is also enriched in  $\rm [Mg/Fe]$ compared to the thin disk population. Figure~\ref{fig:mgfe} shows that the stars with ages corresponding to the peak around $13$\,Gyr indeed are alpha-enhanced. This abundance difference is consistent with core-collapse SNe enrichment from a massive progenitor population \citep{2011MNRAS.414.2893F}, followed by a star formation gap between 10 to 12\,Gyr (in MIST age scale), indicated by the gap between two densities in the AMR \citep{2008MNRAS.384..173F,2008MNRAS.384.1563F}. This thick disk population has been observed in previously published age distributions \citep[e.g.,][]{2016MNRAS.455..987C,2019arXiv190607489R}, however are sample is much larger and with selection effects properly taken into account.

One must be cautious when claiming sub-populations in the AMR, especially when the weighting is sensitive to assumptions we made when calculating the selection function. We have performed additional tests to verify if the age distribution is in fact double-peaked. In the first test, we check if the GALAH selection function is biased against the thick age peak at 13\,Gyr.  This is done by building a synthetic model using MIST isochrones. We simulate the thin disk age distribution as a uniform function between 0 to 10\,Gyr, the thick disk as a Gaussian centered at 12\,Gyr with $\sigma=$ 2\,Gyr. For $\rm [Fe/H]_{bulk}$ we choose two Gaussians centered at 0.0 and -0.5\,dex with $\sigma=$ 0.2\,dex for thin and thick disks, respectively. We sample the age-metallicity distribution for one million mock stars with masses determined by the \citet{2001MNRAS.322..231K} IMF. Isochrones are used to obtain synthetic stellar parameters for this sample, with magnitudes projected to random distances up to 2000\,pc (as most of our sample is relatively near by). Finally we apply the GALAH selection cuts described in Section~\ref{sec:selection}. Indeed we find the secondary thick disk age peak to be greatly reduced after such a GALAH target selection. The aim of our simple model is not to model the thin/thick disk dichotomy in the most accurate way, but to illustrate how the selection function can potentially remove the presence of the thick disk age peak in the raw data. 

In the second test, we attempt to 'deconvolve' the age distribution, given the uncertainties in the stellar parameters and see if the observed age distribution can be more adequately described using a double peak model compared to a single peak model. We assume that the intrinsic age distribution can be described either by a single Gaussian or by two Gaussian distributions of different mean age, standard deviation and relative amplitude. These are convolved with the expected age uncertainties using a mock sample. The convolved distribution is then fit using the single/double peak Gaussian and the likelihood of the observed age distribution calculated. The two-peak model does provide a better fit, however neither is overly well-fitted, an indication that the intrinsic age distribution being inherently non-Gaussian. This is confirmed from a Kolmogorov-Smirnov test \citep{massey1951kolmogorov} on the samples, which both give very low p-values, i.e. the observed distribution is very likely to be different from a Gaussian.

To summarise, these additional tests show that the GALAH selection function has the effect of reducing the signal of the chemical thick disk in our age distribution. However the intrinsic age distribution is likely not inherently Gaussian, which prevents us from  making a definite conclusion of the existence of the secondary peak and its connection to the local chemical thick disk, despite its enticing implications. A more in depth treatment of the proposed secondary peak would be fitting the profile with more sophisticated forms informed by galactic models, combined with GALAH DR3 which offers better age resolution. We intend to return to this issue in a future study.

\subsection{Implications for radial migration}

Metallicity scatter in the AMR is a crucial constraint on radial migration, emphasised by e.g., \citet{1993A&A...280..136F} and \citet{2002MNRAS.336..785S}. As the Galaxy formed inside-out, stars near the center are more metal enriched compared to those at larger radii. This essentially translates to a negative metallicity gradient as a function of galactic radius, with distinct metallicities at each radius. Without radial migration, stars at a given radius will follow a tight AMR. Under migration, stars with same age but formed at different radii (hence having different metallicities) can move in and out of the birth neighbourhood, adding scatter to the local AMR. Potential mechanisms of radial migration have been widely discussed in literature \citep[e.g.,][]{1977A&A....60..263W,2002MNRAS.336..785S,2010ApJ...722..112M}. The observed scatter in $\rm [Fe/H]$ for any given age bin in the AMR is expected from Galactic chemo-dynamical evolution models due to the effect of radial migration as argued in the works of e.g.,~\citet{2008ApJ...684L..79R, 2013A&A...558A...9M, 2015A&A...580A.126K, 2017MNRAS.467.1154S,2009MNRAS.396..203S,2018ApJ...865...96F}. The AMR lower/upper boundary is traced by stars born in the outer/inner disk \citep[see for example Figure 4 of][]{2013A&A...558A...9M}.

\begin{figure}
\includegraphics[width=9.2cm]{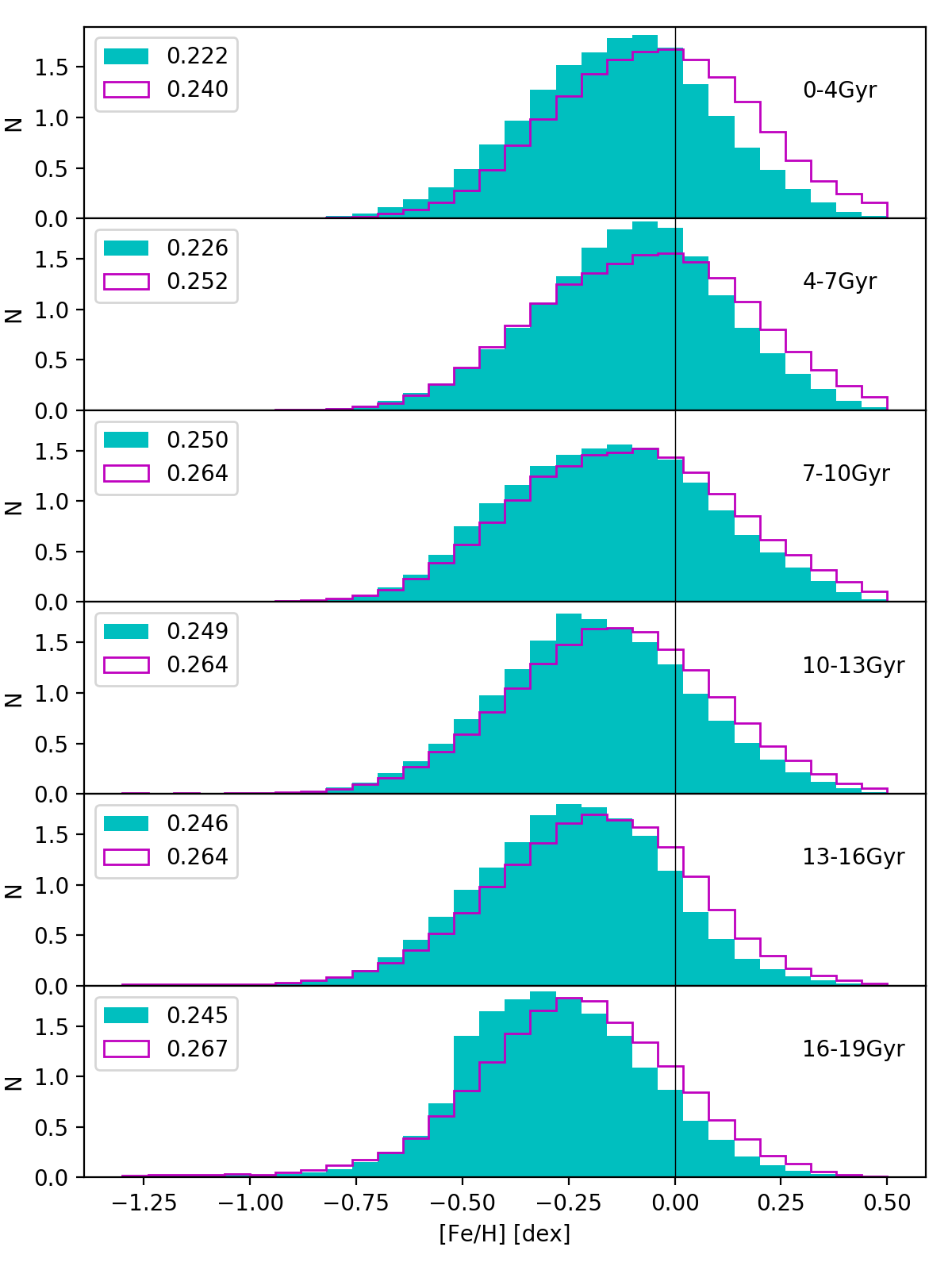}
\caption{Metallicity distribution function for the weighted AMR at 3\,Gyr age bins. Cyan and purple histograms represent bulk and surface metallicities, respectively. The numbers on the left hand side are standard deviations of the distributions. }
\label{fig:stds}
\end{figure} 

Figure~\ref{fig:stds} shows the weighted AMR scatter per age bin for both bulk (cyan) and surface (magenta) metalliticies. Scatter in $\rm [Fe/H]_{bulk}$ is smaller at all ages compared to that of $\rm [Fe/H]_{surf}$. By utilising $\rm [Fe/H]_{bulk}$ instead of $\rm [Fe/H]_{surf}$, we have reduced the effects of stellar evolution on metallicity by directly measuring the initial bulk composition rather than the evolved surface composition \citep{2018MNRAS.477.2966L}. Indeed this is supported by $\rm [Fe/H]_{bulk}$ having smaller scatter than $\rm [Fe/H]_{surf}$ for open clusters, where we would expect all stars to share the same metallicity (with scatter mostly due to measurement). 

In the thin disk, a  mild increase in scatter is observed for the first 3 age bins, reaching its peak at 5\,Gyr (judging from middle panel of Figure~\ref{fig:amr}), indicative of the duration of radial migration reaching its maximum effect. The extent of the MDF spread at young ages can be used to test the efficiency and time scale of radial migration. This being said, we want to stress again that our age scale is expanded due to having no $\rm \alpha$-enhancement, sequence of these events should be taken in a relative context. 

Radial migration predicts an increase in scatter in older ages, which is not observed here, as the scatter is more or less the same for the three oldest bins. This could be due to the uncertainty distributions being elongated in the age direction, effectively squashing the AMR in the said direction. On the other hand, it has been suggested that the thick disk formed from a pool of well mixed materials due its lack of observable radial metallicity gradient \citep[e.g.,][]{2015ApJ...808..132H, 2006ApJ...636..804A}, this will cause a halt in metallicity dispersions of older stars \citep[also observed by][]{2019arXiv190202127D}.

Peaks of the metallicity distribution function for the youngest bin (1-4\,Gyr) appear to be sub-solar and become even more metal-poor toward older age bins.  This could suggest that the Sun migrated from a metal-rich inner region of the Galaxy \citep[e.g.,][]{2018ApJ...865...96F}. The global enrichment history of the Milky Way is reflected by the older bins having more metal-poor stars. For the oldest bin, we find an unsymmetrical distribution, with more metal-poor stars compared to metal-rich stars. The long trail of metal-rich stars (if not all stars within this oldest bin) are likely to have originated from the inner Galaxy. Qualitatively, the GALAH MDFs agree well with those of LAMOST presented in \citet{2018MNRAS.475.3633W} (middle column of their figure 13), including a sub-solar peak for the youngest bins and an unsymmetrical MDF for the oldest bin. 

\begin{figure}
\includegraphics[width=9.2cm]{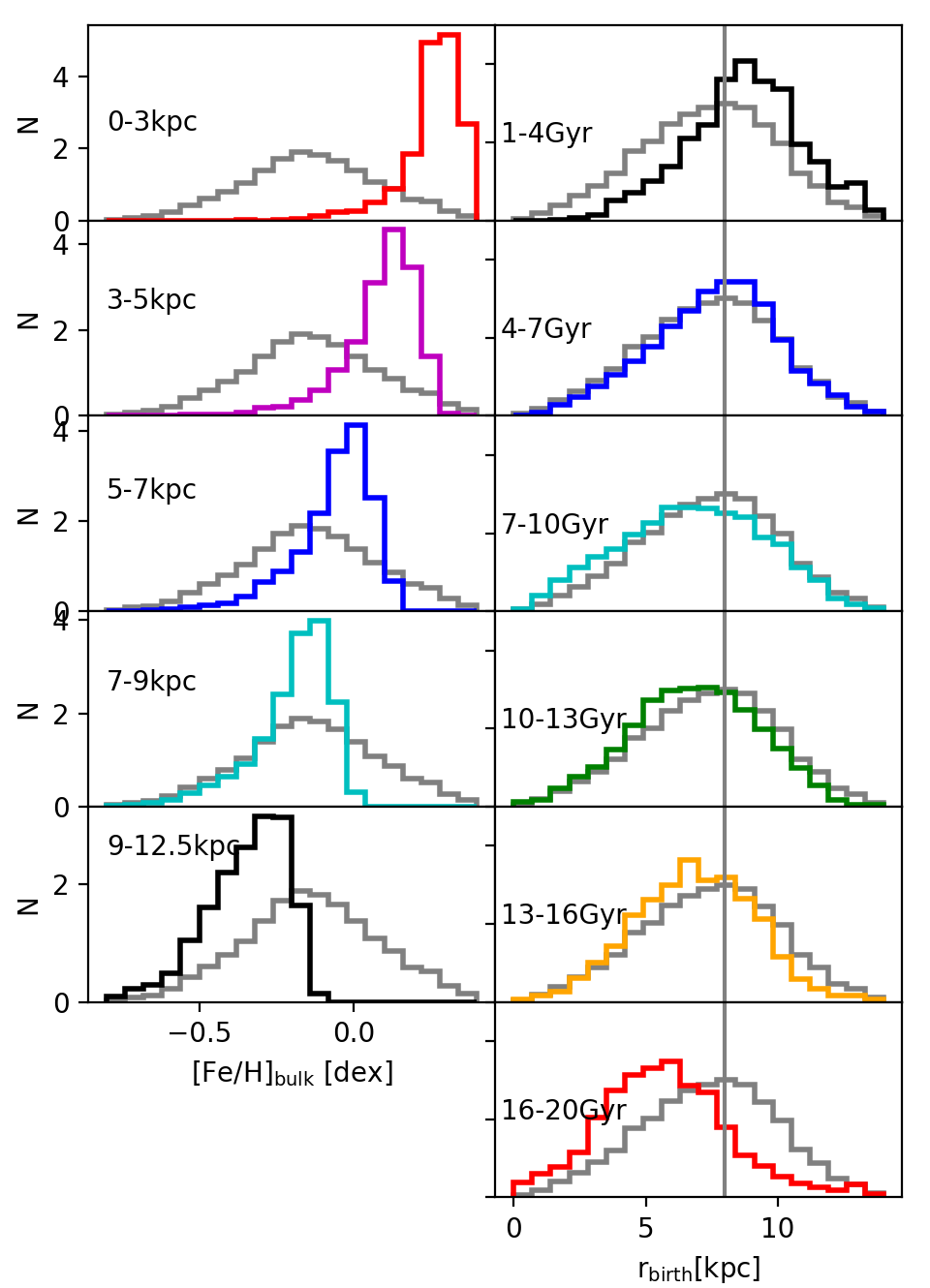}
\caption{Left panels: distribution of $\rm [Fe/H]_{bulk}$ at different $\rm r_{birth}$ intervals, the gray distribution shows the overall $\rm [Fe/H]_{bulk}$ across all bins. Right panels: distribution of $\rm r_{birth}$ at different age intervals, the gray line represents 8\,kpc and the gray distribution is the overall $\rm r_{birth}$ across all ages. }
\label{fig:rbirth}
\end{figure}

To make a quantitative comparison with theory, we calculate birth radii ($\rm r_{birth}$) for a set of 92,414 stars with robust ages (relative age uncertainly under <20\%), based on the methodology presented in \citet{2018MNRAS.481.1645M} and taking selection effects into account. In short, birth radii are derived by projecting stars onto ISM metallicity gradients corresponding to their ages. Both the resulting $\rm r_{birth}$ distribution and ISM metallicity gradients are calibrated on AMBRE:HARPS \citep{2014A&A...570A..68D,2017A&A...608L...1H} and HARPS-GTO \citep{2017A&A...606A..94D,2012A&A...545A..32A} samples. In this paper, we adopt the predicted ISM metallicity gradients presented in \citet{2018MNRAS.481.1645M}. Left panels of Figure~\ref{fig:rbirth} show the distribution of $\rm [Fe/H]_{bulk}$ at different $\rm r_{birth}$ intervals. There is an inverse relationship between $\rm r_{birth}$ and $\rm [Fe/H]_{bulk}$: stars born in the inner galaxy are the most metal rich, with metallicities decreasing at larger radii. Right panels of Figure~\ref{fig:rbirth} shows the distribution of $\rm r_{birth}$ at different age bins. Stars are born inside out, where older stars are more likely to be born in the inner galaxy. We note that the youngest stars have a $\rm r_{birth}$ distribution peaking slightly outside of the solar neighborhood (top right panel), which is most likely due to the ISM metallicity gradients adopted by \citet{2018MNRAS.481.1645M} were calibrated on the HARPS samples, not optimised for GALAH. These two results are in good agreement with overall trends presented in \citet{2018MNRAS.481.1645M}. Comparing to the $\rm [Fe/H]$ distributions at different observed radii bins ($\rm r_{observed}$) presented in \citet{2015ApJ...808..132H}, we find that our distributions do not change skewness across $\rm r_{birth}$. The change in skewness of the \citet{2015ApJ...808..132H} sample is due to radial migration (their figure. 10), by which $\rm r_{birth}$ is not impacted by.

\section{Age-abundance trends} \label{sec:trends}

\begin{figure}
\includegraphics[width=8cm]{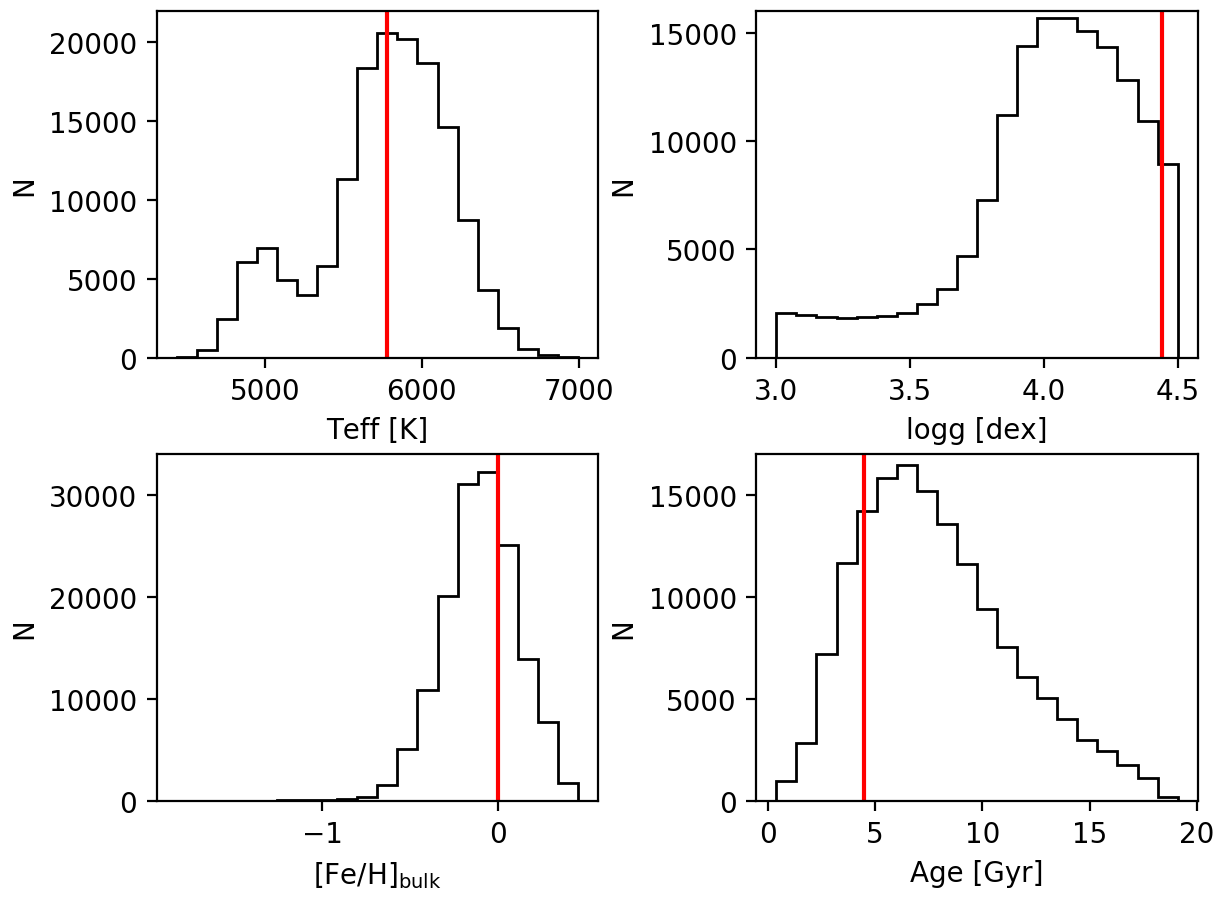}
\caption{ Distribution of stellar parameters for our sample. Red lines indicate solar values.}
\label{fig:dists}
\end{figure}

Here we present the age and abundance trends for our sample, with Figure~\ref{fig:dists} showing the resulting distributions in stellar parameters. The cuts we have made in temperature and gravity should not affect abundance ratio trends with age. On the contrary, by restricting our sample to a limited window of evolutionary stages, we are more confident in their abundances, as any systematic errors in the analysis should at least be in the same direction and of similar magnitude for all stars, making any trends with age easier to discern. Furthermore, as discussed previously, isochrone ages are the most reliable in this region.  

\subsection{Comparison with solar twins}

Solar twin studies are good benchmarks in abundance trend studies due to the extremely high precision achieved in strictly differential analyses relative to the Sun \citep[e.g.,][]{2009ApJ...704L..66M,2018arXiv180202576B,2016MNRAS.463..696L}. Hence it is important to compare the GALAH abundance trends with literature solar twins. Normally, solar twins are defined as stars within 100K in $\rm T_{eff}$, 0.1\.dex in log$\rm g$ and 0.1\.dex in $\rm [Fe/H]_{surf}$ of the solar value \citep[e.g.,][]{2009ApJ...704L..66M}, although this characterisation is somewhat arbitrary.

\begin{figure*}
	\includegraphics[width=160mm]{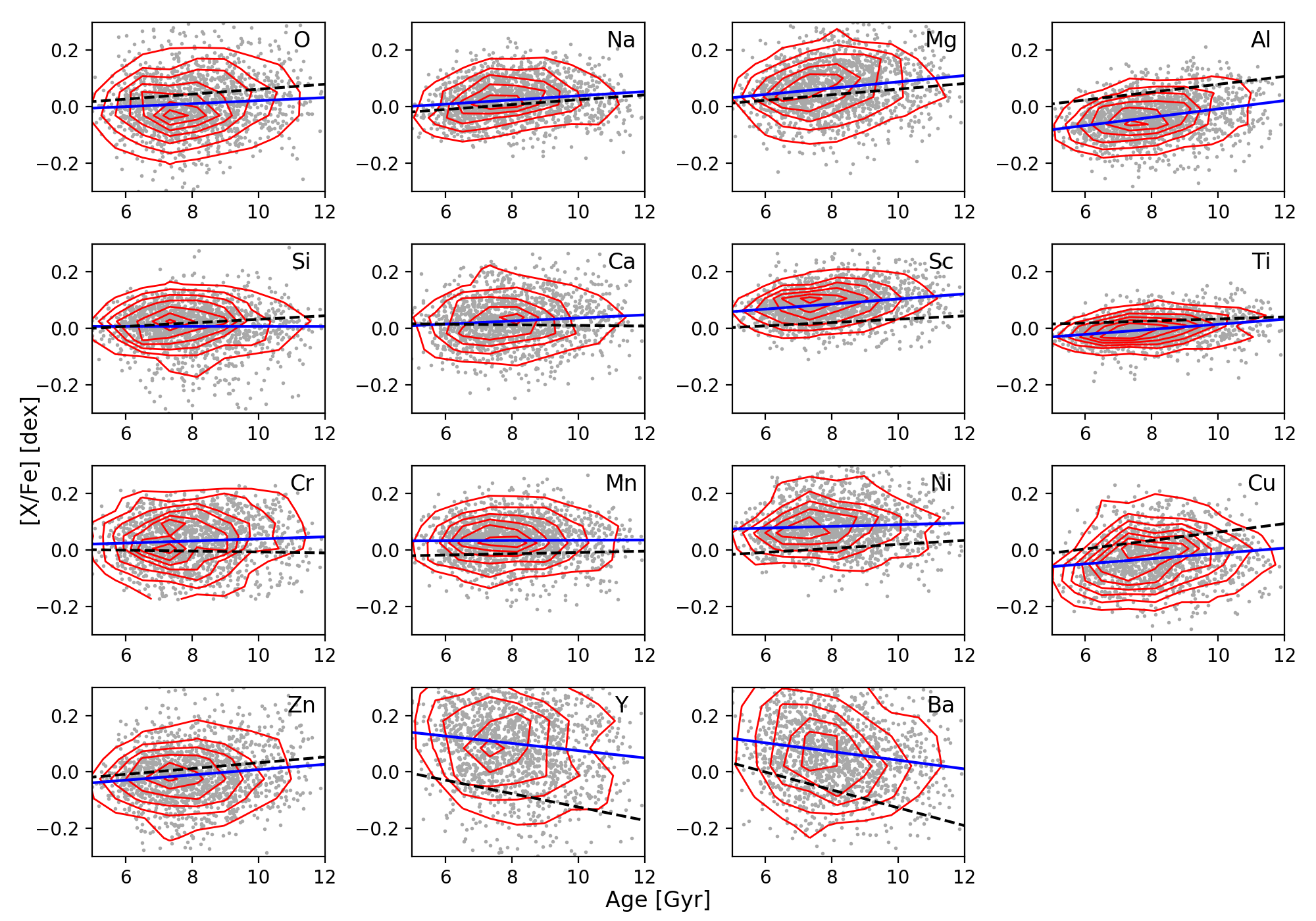}
	\caption{ Age-abundance trends for GALAH solar twins, across 15 elements, the solid blue lines are fits from this work, the black dashed lines are fits from \citet{2018arXiv180202576B}.}
    \label{fig:solart}
\end{figure*}

Figure~\ref{fig:solart} shows the abundance-age trend comparison between our solar twins (1574 stars, with relative uncertainty under 30\%) and solar twins from  \citet{2018arXiv180202576B} (to allow for direct comparison, we employed the same fitting algorithm as \citet{2018arXiv180202576B}). However we must again stress the caveat that the age scales are not directly comparable, as our ages are higher for reasons explained in Section~\ref{sec:aged}. Our sample has few stars below 5\,Gyr, mainly because of the different stellar parameter distributions between \citet{2018arXiv180202576B} and GALAH solar twins- for instance, their sample contains more stars with lower log g (proportionally) than that of GALAH, furthermore, their sample includes stars outside of the solar twin selection box. In addition, our expanded age scale (stars with solar parameters are older in this scheme) and the relative age uncertainty cut also contribute to the paucity of stars below 5\,Gyr.

Overall, our results (blue lines) agree with those presented in \citet{2018arXiv180202576B} (black dashed lines) for solar twins. Some elements have the same slopes, but with slight offsets (e.g., Al, Sc, Mn, Ni, Cu, Y and Ba), which are largely attributed to slightly different methods in determining solar reference abundances (e.g., choice of lines). Compared to  \citet{2018arXiv180202576B}, our Ti and Ca abundances increase slightly more with age, whereas Si is almost flat with age (although an increasing trend with age is observed in the bigger sample, see Figure~\ref{fig:fehb_bins}). 

\subsection{Chemical clocks and solar analogues}

We expand our sample to `solar analogues' to include younger and older stars largely missing from the GALAH solar twin sample. Borrowing the definition from \citet{feltzing2016metallicity}, we take solar-analogues to be stars within 100K of solar $\rm T_{eff}$ (with no restriction on metallicity and gravity). This extended sample allows us to test various chemical clocks found in literature \citep[e.g.,][]{2016A&A...590A..32T, feltzing2016metallicity,2015A&A...579A..52N} using a much larger collection of stars. We restrict our analogues to only those with well determined ages (age uncertainty less than 10\%).

\begin{figure*}
\includegraphics[width=16cm]{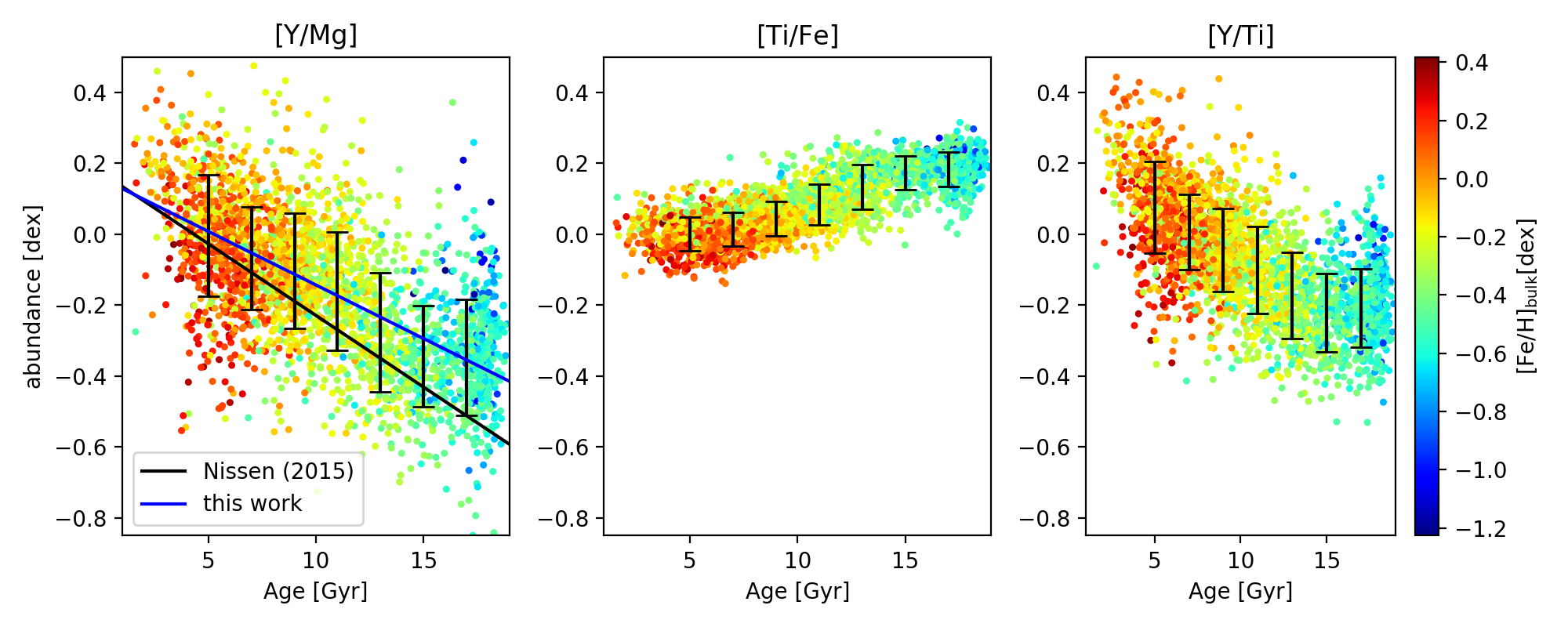}
\caption{ Chemical clocks: $\rm [Y/Mg]$ (left panel), $\rm [Ti/Fe]$ (middle panel) and $\rm [Y/Ti]$ against age (right panel), colour coded in  $\rm [Fe/H]_{bulk}$. Only solar analogues with age uncertainty less than 10\% are included. Over plotted are spreads (standard deviation) in abundance direction at different age bins.}
\label{fig:mgti}
\end{figure*} 

Figure~\ref{fig:mgti} shows the $\rm [Y/Mg]$-age trend for 2713 solar analogues. As expected, a downward trend with age is observed, resulting from the delayed enrichment of AGB products (Y) relative to core-collapse SNe (Mg). We also find intermediate age stars usually have $\rm [Y/Mg]$ and $\rm [Fe/H]$ in sync (i.e. higher $\rm [Y/Mg]$ means higher $\rm [Fe/H]$), as reported in \citet{feltzing2016metallicity} (right panel of their figure 1). For the most metal-rich stars, they have lower $\rm [Y/Mg]$ at a given age. Comparing our trend (blue) with that of \citet{2015A&A...579A..52N} (black), there is a difference in slope, but otherwise the trends are quite similar. It should be noted that the \citet{2015A&A...579A..52N} sample only contains solar twins (within 100\,K of $\rm T_{eff}$, 0.15\,dex in log $\rm g$ and 0.1\,dex in $\rm [Fe/H]_{surf}$ of the solar value), with high precision, differential abundances. 

Figure~\ref{fig:mgti} also shows the $\rm [Ti/Fe]$-age trend for 2741 solar analogues with reliable ages. \citet{feltzing2016metallicity} argues that $\rm [Ti/Fe]$ is not a straightforward age indicator as metallicity introduces a large amount of scatter. Here we show that if restricted to only solar analogues, $\rm [Ti/Fe]$ is a viable age indicator, with smaller scatter in abundance for all ages, but also with a smaller slope in abundance vs age compared to $\rm [Y/Mg]$. Finally, right panel of Figure~\ref{fig:mgti} shows the $\rm [Y/Ti]$-age trend for 2741 solar analogues, with comparable age sensitivity compared to  $\rm [Y/Mg]$ and  $\rm [Ti/Fe]$, and smaller scatters in the abundance direction compared to $\rm [Y/Mg]$. Using this extended sample, we show that the chemical clocks presented in literature are indeed plausible for solar analogues, with $\rm [Y/Ti]$ and $\rm [Ti/Fe]$ being the most age sensitive. 

\subsection{Age-abundance trends by metallicity}

We fit all age-abundance trends using orthogonal distance regression (\texttt{scipy.odr}), allowing the fit to be constrained by uncertainties ($\rm \sigma$) in both directions. We compare this fitting method with that used in \citet{2018arXiv180202576B} which accounts also for intrinsic scatter and find very little difference. We modify the uncertainties in ages to be $\rm \sigma_{modified}=\sigma P $, adding more weight to stars biased against the selection, we note that this weighting is essentially the same Monte Carlo fitting the data, sampling in a way that corrects for the selection biases.  A floor of $\rm P=0.001$ is introduced such that our fits will not be skewed by a small number of stars with extremely low probabilities. We do not modify the uncertainties in abundances because probabilities of a star being observed by GALAH should be independent of $\rm [X/Fe]$ to the first order. 

Stars are binned in four $\rm [Fe/H]_{bulk}$ bins: $\rm [-0.5,-0.1], [-0.1,0], [0,0.1], [0.1,0.5] $ dex, the bin widths are wider for metal-rich and metal-poor bins to allow enough stars in these bins to properly constrain the fit. Here we do not consider uncertainties in metallicity when drawing up the bins, as there should be enough stars per bin such that we are effectively fitting for abundance trends with age, conditioning on $\rm [Fe/H]_{bulk}$, essentially assuming that the uncertainty of [Fe/H] is negligible compared to the age uncertainty. We delineate 12\,Gyr as the thin-thick disk separation and fit each disk separately with a linear function. The choice of 12\,Gyr is somewhat arbitrary; we tested a range of ages around this value and found no significant changes to the trends. 

\begin{figure*}
	\includegraphics[width=160mm]{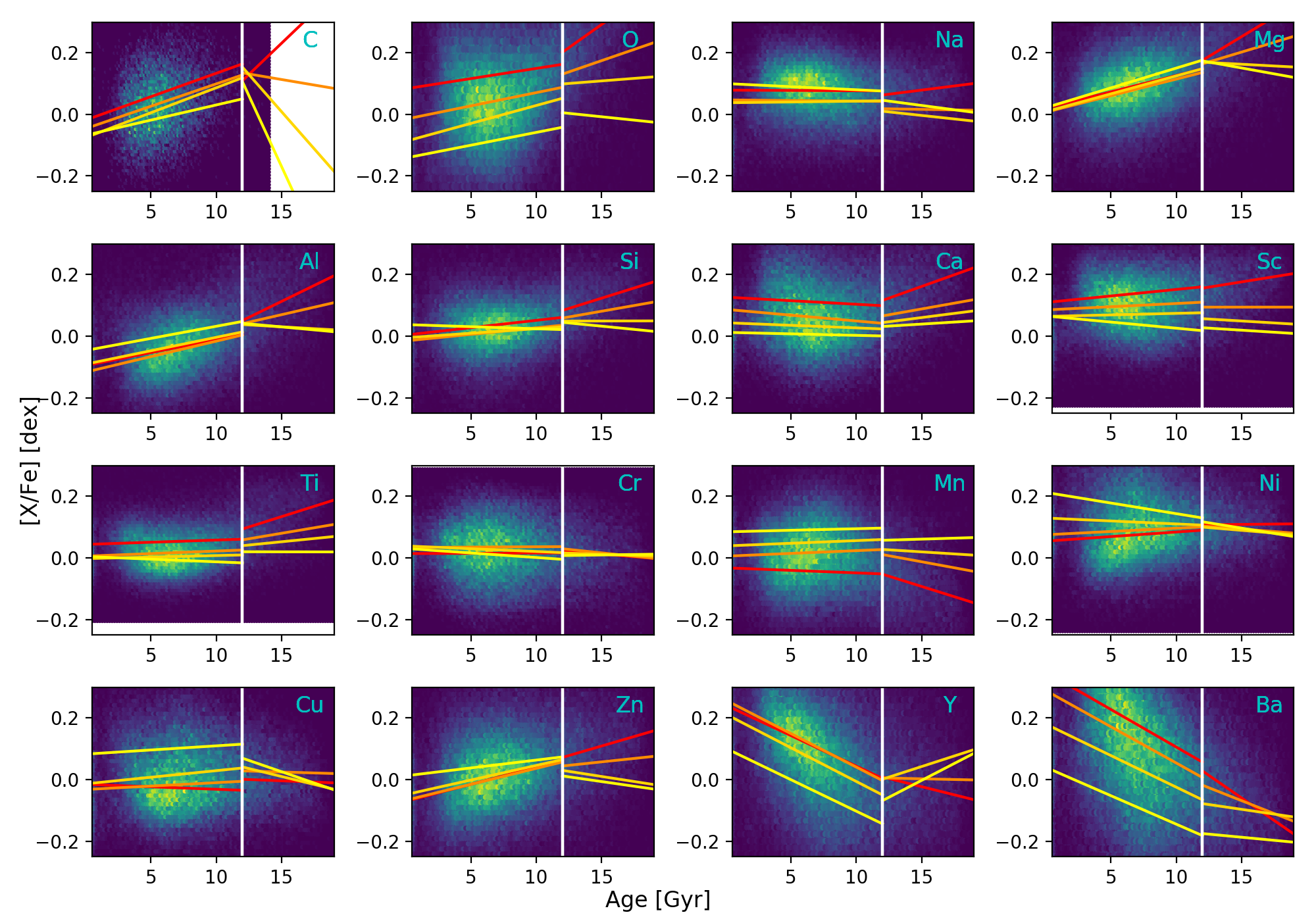}
	\caption{ Age-abundance ratio trends for 16 elements, at four $\rm [Fe/H]_{bulk}$ bins: $\rm [-0.5,-0.1]\ (red), [-0.1,0]\ (orange), [0,0.1]\ (gold), [0.1,0.5]\ (yellow) $, fitted linearly and separately for thin/thick disks (above and below 12\,Gyr).  }
    \label{fig:fehb_bins}
\end{figure*}

Figure~\ref{fig:fehb_bins} shows the trends in different $\rm [Fe/H]_{bulk}$ bins for 16 elements (their intercepts and gradients are listed in Appedix~\ref{sec:appedix}) . Looking at the overall trends, alpha abundances such as O, Si, Ti and Mg have  higher abundances (relative to iron) in the thick disk compared to the thin disk. The abundance ratios are higher for older ages for the thin disk, with the trends steepening with age in the thick disk, \citep[in agreement with previous findings using smaller, but more precise samples, e.g.,][]{2018arXiv180202576B,2015A&A...579A..52N, 2018MNRAS.477.2326F}. This is largely due to their progenitor core-collapse SNe being more dominant in the early Galaxy, increasing alpha abundances before the onset of SNe Ia, producing Fe. The interplay between two classes of SNe is most prominent in Mg and O, reflected by their steeper slopes in the thin disk compared to other alpha elements. Both elements have  largest core-collapse SNe contribution, hence are commonly referred to as the pure alpha elements.

\begin{figure*}
	\includegraphics[width=160mm]{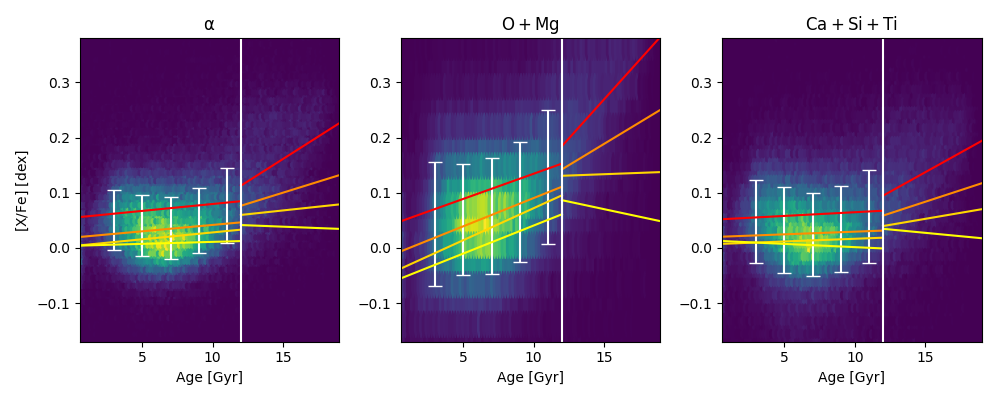}
	\caption{ Left panel: Age-alpha trends at  four $\rm [Fe/H]_{bulk}$ bins: $\rm [-0.5,-0.1]\ (red), [-0.1,0]\ (orange), [0,0.1]\ (gold), [0.1,0.5]\ (yellow) $, fitted linearly and separately for thin/thick disks (above and below 12\,Gyr). Over plotted are the spread (standard deviation) of $\rm [\alpha/Fe]$ at age bins in the thin disk. Middle panel: same as left, but with $\rm [(O+Mg)/Fe]/2$. Right panel: same as left, but with $\rm [(Ca+Si+Ti)/Fe]/3$.}
    \label{fig:afe}
\end{figure*}

In the thin disk, alpha abundance ratio trends have similar gradients for all metallicity bins (i.e. uniform enrichment speeds with respect to iron across all metallicities). But in the thick disk the trends flare out separately, with the most metal-poor bin having the largest increase with age. Certain alpha elements (e.g., O, Mg) have negative slopes for the most metal-rich bin, this could be due to having too few stars with high metallicity and age to properly constrain the fit. Ca behaves differently to other alpha elements, in the sense that its abundance ratio decreases with age in the thin disk.  \citet{2019arXiv190202127D} speculates this oddity is due to enrichment from a new SNe subclass: calcium-rich transients \citep{2010Natur.465..322P}. We also note the solar metallicity trends of Ca (orange and gold lines) behave slightly differently compared to those of solar twins in Figure~\ref{fig:solart} These differences can be attributed largely to our relaxed selection criteria here (all stars within 0.1\,dex of solar metallicty are selected regardless of gravity and temperature.)

Al and Zn behave similarly to the alpha elements in both the thin (abundance ratios increase with age)  and thick (trends flaring out) disks. Both elements trace alpha behaviour as both are produced by core-collapse SNe/hypernovae \citep[e.g.,][]{2013ARA&A..51..457N,2017ApJ...835..224A}, with their abundances decreasing over time due to the onset of SNe Ia. Figure~\ref{fig:afe} shows the spread in abundance per age bin for three different abundance combinations in the thin disk: average of all alpha elements (left panel), O$+$Mg (middle panel) and Ca$+$Si$+$Ti (right panel). Average alpha gives smaller spread compared to other two combinations, with the spread being reduced by combining multiple elements together. Pure alpha elements O$+$Mg have the steepest trends, but also larger scatter in abundance ratio at any given age, which is mainly introduced by observational uncertainties in O. These steep trends are in contrast with the flat trends in Ca$+$Si$+$Ti, which are all alpha elements with contamination from non core-collapse SNe sources. 

\begin{figure}
\includegraphics[width=9.2cm]{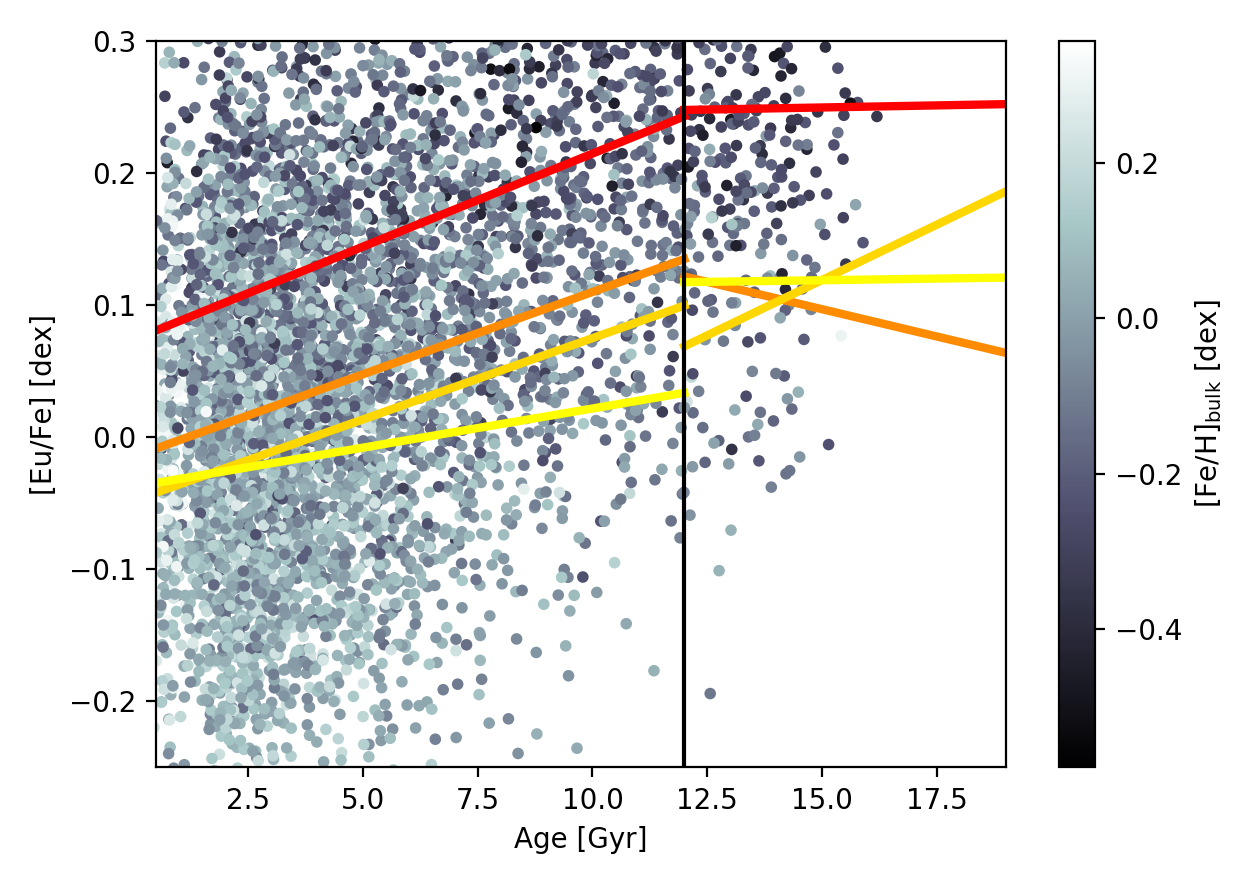}
\caption{$\rm [Eu/Fe]$ trends over time, across four $\rm [Fe/H]_{bulk}$ bins: $\rm [-0.5,-0.1]\ (red), [-0.1,0]\ (orange), [0,0.1]\ (gold), [0.1,0.5]\ (yellow) $, fitted linearly and separately for thin/thick disks (above and below 12\,Gyr). Points are colour-coded according to  $\rm [Fe/H]_{bulk}$. We do not have enough stars with good $\rm [Eu/Fe]$ abundances to make a density plot similar to other elements in Figure~\ref{fig:fehb_bins}. Intercepts and gradients of the thin-disk trends are listed in Appendix~\ref{sec:appedix}.}
\label{fig:eufe}
\end{figure}

On the other hand, s-process elements are produced in low-mass AGB stars, hence we expect their abundance ratios to increase over time, this is observed in both Y and Ba in the thin disk, where Y has smaller metallicity scatter than Ba. A gradient reversal is observed for both elements in the thick disk \citep[similar to e.g.,][]{2018A&A...617A.106M,2019arXiv190202127D,2018MNRAS.474.2580S}. Y (a light s-element) has higher abundance ratios compared to Ba (a heavy-s element) at a given age in the thick disk, this behaviour is also observed by \citet{2019arXiv190202127D}. Cu also exhibits a slight gradient reversal between thin and thick disks, with abundance ratio increase with age in the thin disk. Lastly, Eu (a r-process element) shows an increase in abundance ratio at higher ages in the thin disk, in contradiction with the other two s-process elements Y and Ba. Some studies have advocated the main production site for r-process elements are neuron star mergers \citep[e.g.,][]{freiburghaus1999r,wanajo2014production}. This should mean that [Eu/Fe] has time-delay features similar to Y and Ba, which is not observed in our data (Figure~\ref{fig:eufe}). It has been suggested that production channels with short time-delays have played a role in producing Eu, in addition to neutron star mergers, such as core-collapse SNe \citep[e.g.,][]{2012MNRAS.421.1231T}.

Over all, our trends in the thin disk over metallicity bins agree well with those presented in  \citet{2019arXiv190202127D} in terms of both gradients and order of individual metallicity bins. This is remarkable because \citet{2019arXiv190202127D} is restricted in surface metallicities and has a small sample size compared to this study. Additionally, they have a much higher resolution ($R\sim 115,000$), compared to our resolution of 28,000. \citet{2019A&A...622A..59T} shows at least some scatter in $\rm [Y/Mg]$-age relation could be contributed by different chemical evolutions undergone by thick and thin disks. This is done by fitting age-abundance trends for low and high alpha sequences which shows distinct trends for these two populations. Here we employ age as the thin-thick distinction and also find separate trends for many more elements (e.g., Y, Ca and Ti). Furthermore, by fitting over metallicity bins, we find some elements to be more metallicity dependent than others. Mg, Al and Zn have similar and tight abundance ratio trends across all metallicity bins, hence are good chemical clock candidates. Other elements such as Mn and Ni have high scatter and different slopes in their abundance ratio trends, and therefore are poor candidates for chemical age estimations. This is perhaps not surprising because Mn and Ni are expected to be produced in lock-step with [Fe/H] (we note that Mn has a nearly flat [Mn/Fe] vs [Fe/H] trend when taking into account non-LTE effects in the spectral line formation \citep[][]{2008A&A...492..823B}). Ideally, for a good chemical clock the two elements should be produced by two distinct channels that have different yield ratios and delay times, in which case the abundance ratio translates into an absolute time-scale.

\section{Conclusions and discussions}

In this study we present isochrone ages and initial bulk metallicities ($\rm [Fe/H]_{bulk}$) of 163,722 stars from the GALAH survey. The stars are mainly located on the turn-off and subgiant regions of the HRD ($\rm 7000K>T_{eff}>4000K$\ and $\rm log g>3$\,dex), which allow us to obtain robust isochrone ages for them, with $\sim$ 80\% of ages having relative uncertainty less than 30\%. One caveat in using MIST isochrones is that due to different solar abundances, lack of alpha enrichment and potential abundance scale mismatches between MIST and GALAH, our ages scale is more extended compared to that of literature. Further more, uncertainties in stellar parameters are also contributing factors for older ages among some stars. Hence it is important to consider everything in a relative chronology, instead of absolute ages. 

We employ a simple method of modelling the selection effects introduced by the survey selection function, as well as limitations in the stellar parameter pipeline. Using the resulting observational probabilities, we are able to correct for the selection effect on the AMR. Without correction, we obtain an AMR similar to that which is widely reported in literature: flat, with large scatter in metallicity for all ages before $\sim$12 Gyr, then a slight downward trend for the oldest stars. With correction, we find the age distribution displays a secondary peak near 13\,Gyr. One possible explanation is that this secondary peak is a manifestation of the thick disk, as its constituent stars are older, more metal-poor and high in $\rm [Mg/Fe]$. This is consistent with a two phase star formation scenario, one for each disk \citep{1998A&A...338..161F}, also found in smaller studies \citep[e.g.,][]{2016MNRAS.455..987C,2019arXiv190607489R}. However we caution that the secondary peak is on the threshold of detection. It is possible to verify this observation by performing a more statistically robust inference by constructing a hierarchical model which folds in both the observed data and GALAH selection cuts. A potential follow up of this study would be to combine such a model with the next GALAH data release, which will encompass more stars and more robust parameters/abundances. Using this model, it is possible to forward model the observed distribution by including also the target selection effect to constrain the intrinsic AMR, allowing us to determine if the double peaked solution is the only unique solution which fits the observed data.

We also find the GALAH abundance-age trends for solar twins to be compatible with the high precision differential analysis literature \citep{2018arXiv180202576B}. Analysing the abundance-age trends of different $\rm [Fe/H]_{bulk}$ bins we find most trends behaving according to stellar nucleosynthesis predictions. Our trends agree with trends in surface metallicity ($\rm [Fe/H]_{surf}$)-bins presented in \citet{2019arXiv190202127D}, despite the differences in sample resolution and size. Our Eu trends show a lack of time-delay features, indicative of non-neutron merger contribution. However we caution our Eu measurements are few compared to other elements, hence GALAH DR3 will be a good sample to test this conclusion.

\section*{Acknowledgements}

YST is supported by the NASA Hubble Fellowship grant HST-HF2-51425.001 awarded by the Space Telescope Science Institute.  LC is the recipient of the ARC Future Fellowship FT160100402. This research was partly supported by the Australian Research Council Centre of Excellence for All Sky Astrophysics in 3 Dimensions (ASTRO 3D), through project number CE170100013. SLM acknowledges funding from the Australian Research Council through Discovery grant DP180101791, and from the UNSW Scientia Fellowship program. JDS acknowledges funding from the Australian Research Council through Discovery grant DP180101791. M{\v Z} acknowledges funding from the Australian Research Council (grant DP170102233). JK, TZ and KC acknowledge financial support of the Slovenian Research Agency (research core funding No. P1-0188).




\bibliographystyle{mnras}
\bibliography{mnras_template} 



\appendix

\section{Gradients and intercepts of abundance trends}\label{sec:appedix}
\begin{table*}
\caption{Gradient (m) and intercept (b) of abundance trends over four bulk metallicity bins in the thin disk ($<$12\,Gyr)}
\begin{tabular}{ccccc}
element & [-0.5,-0.1] m,b & [-0.1,0] m,b  & [0,0.1] m,b & [0.1,0.5] m,b \\
\hline
C       & 0.0152, -0.0185        & 0.0145, -0.0462        & 0.0162, -0.0757        & 0.0098, -0.0676        \\
O       & 0.0066, 0.0832         & 0.0086, -0.0154        & 0.0117, -0.0875        & 0.0083, -0.1416        \\
Na      & -0.0002, 0.0788         & -0.0004, 0.0471         & 0.0005, 0.0378         & -0.002, 0.1002         \\
Mg      & 0.0106, 0.0200           & 0.0108, 0.0069         & 0.0117, 0.0086         & 0.0129, 0.0217         \\
Al      & 0.0084, -0.0956        & 0.0101, -0.1169        & 0.0087, -0.0911        & 0.0079, -0.0467        \\
Si      & 0.0048, 0.0035         & 0.0043, -0.0156        & 0.003, -0.0053        & -0.0014, 0.0375         \\
Ca      & -0.0023, 0.1267         & -0.0038, 0.0870          & -0.0017, 0.0435         & -0.001, 0.0123         \\
Sc      & 0.0042, 0.1092         & 0.0021, 0.0850          & 0.0010, 0.0639         & -0.0040, 0.0656         \\
Ti      & 0.0014, 0.0436         & 0.0017, 0.0055         & 0.0010, -0.0025        & -0.0017, 0.0044         \\
Cr      & 0.0002, 0.0138         & 0.0002, 0.0337         & -0.0018, 0.0386         & -0.0029, 0.0305         \\
Mn      & -0.0016, -0.0324        & 0.0018, 0.0054         & 0.0017, 0.0387         & 0.0010, 0.0844         \\
Ni      & 0.0030, 0.0542         & 0.0021, 0.0749         & -0.0019, 0.1294         & -0.0069, 0.2126         \\
Cu      & -0.0016, -0.0154        & 0.0022, -0.0323        & 0.0043, -0.0139        & 0.0027, 0.0826         \\
Zn      & 0.0111, -0.0708        & 0.0104, -0.0673        & 0.0095, -0.0490         & 0.0051, 0.0122         \\
Y       & -0.0198, 0.2421         & -0.0218, 0.2582         & -0.0219, 0.2128         & -0.0205, 0.1019         \\
Ba      & -0.0244, 0.3497         & -0.0236, 0.2894         & -0.0205, 0.1807         & -0.0185, 0.0401    \\
Eu      & 0.0141, 0.0737        & 0.0125, -0.0148 & 0.0123, -0.0482 & 0.0059, -0.0376
\end{tabular}
\end{table*}

\begin{table*}
\caption{Gradient (m) and intercept (b) of abundance trends over four bulk metallicity bins in the thick disk ($>$12\,Gyr), intercepts are based on the 12\,Gyr demarcation, as shown in Figure~\ref{fig:fehb_bins} }
\begin{tabular}{ccccccccc}
element & [-0.5,-0.1] m,b & [-0.1,0] m,b  & [0,0.1] m,b & [0.1,0.5] m,b \\
\hline
C       & 0.0406, 0.1096          & -0.0072, 0.1348          & -0.0483, 0.1534          & -0.0921, 0.1062          \\
O       & 0.0301, 0.2019          & 0.0146, 0.1308          & 0.0033, 0.0989          & -0.0044, 0.0055          \\
Na      & 0.0054, 0.0629          & -0.0007, 0.0184          & -0.0046, 0.0104          & -0.0056, 0.0459          \\
Mg      & 0.0257, 0.1756          & 0.0128, 0.1642          & -0.0023, 0.1697          & -0.0077, 0.1738          \\
Al      & 0.0208, 0.0507          & 0.0097 , 0.0405          & -0.0024, 0.0376          & -0.0039, 0.0417          \\
Si      & 0.0133, 0.0834          & 0.0074, 0.0586          & 0.0001, 0.0489          & -0.0042, 0.0448          \\
Ca      & 0.0152, 0.1159          & 0.0075, 0.0656          & 0.0054, 0.0440           & 0.0026, 0.0313          \\
Sc      & 0.0068, 0.1558          & 0.0001, 0.0942          & -0.0025, 0.0567          & -0.0027, 0.0272          \\
Ti      & 0.0134, 0.0937          & 0.0072, 0.0584          & 0.0042, 0.0396          & -0.0001, 0.0198          \\
Cr      & -0.0036, 0.0232          & -0.0044, 0.0295          & -0.0015, 0.0158          & 0.0006, 0.0074          \\
Mn      & -0.0133, -0.0531         & -0.0078, 0.0107          & -0.0027, 0.0273          & 0.0013, 0.0570           \\
Ni      & 0.0004, 0.1073          & -0.0033, 0.1035          & -0.0036, 0.1000             & -0.0069, 0.1171          \\
Cu      & -0.0017, 0.0010           & -0.0012, 0.0282          & -0.0103, 0.0410           & -0.0147, 0.0694          \\
Zn      & 0.0124, 0.0719          & 0.0045, 0.0441          & -0.0066, 0.0299          & -0.0061, 0.0118          \\
Y       & -0.0106, 0.0086          & -0.0009, 0.0052          & 0.0137, 0.0012          & 0.0220, -0.0685         \\
Ba      & -0.0293, 0.0280           & -0.0168, -0.0183         & -0.0062, -0.0781         & -0.0040, -0.1753        
\end{tabular}
\end{table*}



\bsp	
\label{lastpage}
\end{document}